\documentclass[11pt,a4paper]{article}
\pdfoutput=1

\usepackage{jheppub}

\usepackage{amssymb}
\usepackage{amsfonts}
\usepackage{graphicx}
\usepackage{epstopdf}
\usepackage{dcolumn}
\usepackage{amsmath}
\usepackage{latexsym,bm}
\usepackage{amsthm}
\usepackage{slashed}
\usepackage{float}
\usepackage{color}
\usepackage{url}
\usepackage{longtable}
\usepackage{subfig}
\usepackage{tikz}
\usepackage[all,cmtip]{xy}

\usepackage{multirow}
\usepackage{longtable}

\usepackage[titletoc]{appendix}
\makeatletter
\newtheorem*{rep@theorem}{\rep@title}
\newcommand{\newreptheorem}[2]{%
\newenvironment{rep#1}[1]{%
 \def\rep@title{#2 \ref{##1}}%
 \begin{rep@theorem}}%
 {\end{rep@theorem}}}
\makeatother

\newreptheorem{lemma}{Lemma}

\newreptheorem{conj}{Conjecture}

\theoremstyle{definition}

\newcommand \xoverline[2][0.75]{
    \sbox{\myboxA}{$\m@th#2$}
    \setbox\myboxB\null
    \ht\myboxB=\ht\myboxA
    \dp\myboxB=\dp\myboxA
    \wd\myboxB=#1\wd\myboxA
    \sbox\myboxB{$\m@th\overline{\copy\myboxB}$}
    \setlength\mylenA{\the\wd\myboxA}
    \addtolength\mylenA{-\the\wd\myboxB}
    \ifdim\wd\myboxB<\wd\myboxA
       \rlap{\hskip 0.5\mylenA\usebox\myboxB}{\usebox\myboxA}%
    \else
        \hskip -0.5\mylenA\rlap{\usebox\myboxA}{\hskip 0.5\mylenA\usebox\myboxB}%
    \fi}
\makeatother

\newcommand{\ba}{\begin{aligned}}
\newcommand{\ea}{\end{aligned}}
\def\be{\begin{equation}}
\def\ee{\end{equation}}
\def\bsp{\begin{split}}
\def\esp{\end{split}}
\def\bea{\begin{eqnarray}}
\def\eea{\end{eqnarray}}

\def \bp{\begin{pmatrix}}
\def\ep{\end{pmatrix}}

\def\R{\mathbb{R}}
\def\N{\mathcal{N}}

\def\Z{\mathbb{Z}}

\def\mk{\mathfrak}

\usepackage{tikz}
\usepackage{diagbox}
\usetikzlibrary{positioning}
\usetikzlibrary{calc}
\usetikzlibrary{decorations.pathreplacing,calligraphy}

\usepackage{xstring}
\usetikzlibrary{decorations.pathmorphing} 
\usetikzlibrary{decorations.markings} 
\usetikzlibrary{snakes}
\usetikzlibrary{arrows} 
\usetikzlibrary{shapes} 
\usetikzlibrary{matrix} 
\usetikzlibrary{positioning} 
\usepackage[english]{babel} 
\usepackage[autostyle]{csquotes}

\usepackage{tikz}
\usetikzlibrary{arrows}
\usetikzlibrary{arrows.meta}
\usetikzlibrary{shapes.geometric,calc,arrows, positioning,shapes.misc,decorations.markings}
\tikzset{
  big arrow/.style={
    decoration={markings,mark=at position 1 with {\arrow[scale=2,#1]{>}}},
    postaction={decorate},
    shorten >=0.4pt},
  big arrow/.default=black}
  
\pgfdeclarelayer{edgelayer}
\pgfdeclarelayer{nodelayer}
\pgfsetlayers{edgelayer,nodelayer,main} 
\tikzstyle{none}=[inner sep=0pt] 

\tikzstyle{NodeCross}=[draw, shape=circle, cross out, inner sep=0pt, minimum size=6pt,line width=0.25mm]
\tikzstyle{Circle}=[draw, shape=circle, black,  fill=black, inner sep=0pt, minimum size=6pt]
\tikzstyle{Star}=[draw, shape=star, fill=black, star points=8, inner sep=0pt, minimum size=8pt]

\tikzstyle{DashedLine}=[-, densely dashed, line width=0.25mm]
\tikzstyle{DottedLine}=[-, dotted, line width=0.25mm]
\tikzstyle{ThickLine}=[-, line width=0.25mm]
\tikzstyle{ArrowLineRight}=[-, -{Stealth[scale=1.75]}, line width=0.1mm, scale=5]
\tikzstyle{RedLine}=[-, draw={rgb,255: red,191; green,0; blue,0}, fill=none, line width=0.25mm]
\tikzstyle{DottedRed}=[-, dotted, draw={rgb,255: red,191; green,0; blue,0}, fill=none, line width=0.25mm]
\tikzstyle{DashedLineThin}=[-, densely dashed, line width=0.125mm, fill=none, draw=black]
\tikzstyle{ArrowLineRed}=[-, -{Stealth[scale=1.75]}, draw={rgb,255: red,191; green,0; blue,0}, line width=0.1mm, scale=5]
\tikzstyle{brane}=[draw]
\tikzset{D7/.style={circle, draw=black, inner sep=0pt, fill=white, minimum size=3mm}}
\tikzset{hasse/.style={circle, fill,inner sep=2pt}}
\tikzset{flavor/.style={regular polygon,fill=white,regular polygon sides=4,inner sep=2.5pt, draw}}
\tikzset{gauge/.style={circle, draw,inner sep=2.5pt}}
\tikzset{gaugeb/.style={circle, draw,fill=black,inner sep=2.5pt}}
\tikzset{gauger/.style={circle, draw,fill=cyan,inner sep=2.5pt}}
\tikzset{gaugeg/.style={circle, draw,fill=red,inner sep=2.5pt}}
\tikzset{SUd/.style={circle, draw=black, inner sep=0pt, fill=yellow, minimum size=2mm}}
\tikzset{bd/.style={circle, draw=black, inner sep=0pt, fill=black, minimum size=2mm}}
\tikzset{wd/.style={circle, draw=black, inner sep=0pt, fill=white, minimum size=2mm}}
\tikzset{Dynkin/.style={circle, draw=black, inner sep=0pt, fill=white, minimum size=2mm}}
\tikzstyle{ligne}=[draw, thick] 
\tikzset{doublearrow/.style={ draw=black!75, color=black!75, thick, double distance=3pt, }} 

\usepackage{mathrsfs}

\usepackage{amsmath}

\newcommand\restr[2]{{
  \left.\kern-\nulldelimiterspace 
  #1 
  \littletaller 
  \right|_{#2} 
  }}

\newcommand{\littletaller}{\mathchoice{\vphantom{\big|}}{}{}{}}

\usetikzlibrary{positioning}

\usepackage{tikz-cd}

\usepackage{tikz}
\usepackage{tikz-3dplot}

\usepackage{enumitem} 

\usepackage{physics}

\usepackage{chemfig}
\usepackage{import}

\makeatletter
\newcommand\xleftrightarrow[2][]{%
  \ext@arrow 9999{\longleftrightarrowfill@}{#1}{#2}}
\newcommand\longleftrightarrowfill@{%
  \arrowfill@\leftarrow\relbar\rightarrow}
\makeatother

\usepackage{booktabs}

\usepackage{braket}

\graphicspath{ {figs/} }

\title{Modified instanton sum and 4-group structure in 4d $\N=1$ $SU(M)$ SYM from holography}

\preprint{\today \hspace*{0.1in} }

\author[\,\clubsuit]{Marwan Najjar}
\emailAdd{marwan.najjar@pku.edu.cn}

\affiliation[\clubsuit]{Center for High Energy Physics, Peking University, \\
Beijing 100871, China.}

\abstract{We study the decomposition of the holographic 4d $\mathcal{N}=1$ $SU(M)$ gauge theory with in the Klebanov-Strassler set-up. In particular, we propose a consistent framework for defining a modified instanton sum and a 4-group structure for the SYM theory, derived from its $AdS/CFT$ construction. To achieve this, we analyze symmetry topological operators associated with continuous $(-1)$-form symmetries, derive the corresponding 5-dimensional Symmetry Topological Field Theory (SymTFT), and impose specific discrete gaugings.}

\begin{document}

\maketitle

\section{Introduction and summary}

Generalized symmetries reformulate the conventional notion of global symmetries in terms of co-dimension one topological operators acting on local operators of quantum field theories (QFTs) \cite{Gaiotto:2014kfa}. Once we adapt to the topological framework, we can consider topological symmetry operators of various co-dimensions. In particular, this defines the notion of $p$-form symmetry, in which the charged defect is given by a heavy non-dynamical $p$-dimensional operator, with a symmetry topological operator of co-dimension $(p+1)$.

Since the pioneering work of \cite{Gaiotto:2014kfa}, significant progress has been made in understanding $p$-form and generalized symmetries of $d$-dimensional field theories. For a sense of the growing body of research, we highlight \cite{Kapustin:2014gua,Gaiotto:2017yup, Gaiotto:2017tne, Cordova:2019uob,Gaiotto:2020iye,Freed:2022qnc,Vandermeulen:2022edk,Wang:2023iqt,Brennan:2024fgj,Antinucci:2024zjp,Bonetti:2024cjk,Apruzzi:2024htg,Aloni:2024jpb,Santilli:2024dyz,Closset:2024sle,Closset:2025lqt,Jia:2025jmn}. From the perspective of string theory and M-theory, the study of symmetries and $p$-form symmetries in QFTs—whether constructed via geometric engineering or the $AdS/CFT$ correspondence—has garnered considerable interest, for instance, \cite{Apruzzi:2019opn,Eckhard:2020jyr,Bah:2019rgq,Acharya:2021jsp,Tian:2021cif,Kim:2021fxx,Apruzzi:2021nmk,Apruzzi:2021phx,Najjar:2022eci,vanBeest:2022fss,Heckman:2022xgu,Apruzzi:2022rei,Najjar:2023hee,Apruzzi:2023uma,Bah:2023ymy,Tian:2024dgl,DelZotto:2024tae,Najjar:2024vmm,GarciaEtxebarria:2024fuk,Gagliano:2024off,Liu:2024znj,Heckman:2024oot,Najjar:2025rgt,Cvetic:2025kdn}. Finally, recent notes and reviews on this rapidly evolving field include \cite{Schafer-Nameki:2023jdn, Brennan:2023mmt, Luo:2023ive, Bhardwaj:2023kri, Shao:2023gho,Iqbal:2024pee}.

Global symmetries provide a framework for defining order parameters that characterize the infrared (IR) behaviour of specific ultraviolet (UV) theories \cite{tHooft:1979rat,Callan:1984sa}. The concept of generalized global symmetry extends this framework, enabling the formulation of a generalized Landau paradigm, e.g., \cite{Chatterjee:2022tyg,Moradi:2022lqp,Wen:2023otf,Bhardwaj:2023bbf,Bhardwaj:2023fca,Bhardwaj:2024qrf}. This paradigm classifies phases of QFTs based on the behaviour of charged defects of $p$-form symmetries treated as order parameters.

The aim of this note is to study the modified instanton sum and the 4-group structure, as introduced in \cite{Seiberg:2010qd,Tanizaki:2019rbk}, for holographic 4d $\mathcal{N}=1$ $SU(M)$ gauge theories \cite{Klebanov:2000hb}. This work is motivated by our previous results in \cite{Najjar:2024vmm}. Central to this work is the presence of 0-dimensional local topological operators, which serve as order parameters in QFTs with a discrete (invertible) $(d-1)$-form symmetry. These operators are closely tied to the concept of decomposition, first explored in \cite{Pantev:2005rh,Pantev:2005wj,Pantev:2005zs,Hellerman:2006zs} and further developed in subsequent works, including \cite{Seiberg:2010qd,Tachikawa:2013hya,Sharpe:2014tca,Tanizaki:2019rbk,Cherman:2020cvw,Cherman:2021nox,Nguyen:2021naa,Sharpe:2022ene,Najjar:2024vmm,Yu:2024jtk}.

\paragraph{Outline of the paper.} The structure of this paper is as follows:
\begin{itemize}
    \item In Section \ref{sec:general-discussion}, we review the construction of the Symmetry Topological Field Theory (SymTFT) within the framework of $AdS/CFT$. We discuss the stringy origin of charged defects and symmetry topological operators associated with $p$-form symmetries. Furthermore, we argue that the consistent and appropriate fluxbrane topological action coincides with the Page charges of the (dual) branes of interest.

    \item In section \ref{sec:(-1)-form-KS-theory}, we analyse the continuous $(-1)$-form and 2-form symmetries of the holographic 4d $\N=1$ $SU(M)$ gauge theory. We derive the corresponding 5-dimensional SymTFT and demonstrate that a modified instanton sum and a 4-group structure can be consistently defined for the gauge theory. This establishes the decomposition of the gauge theory within the $AdS/CFT$ framework.
\end{itemize}

\section{Generalities: SymTFT, defects, and symmetry operators}\label{sec:general-discussion}

In this section, we explore the realization of the SymTFT within the context of $AdS/CFT$ correspondence. We discuss charged defects and symmetry topological operators, with a particular focus on formulating the symmetry topological operators of continuous symmetries as Page charges and revealing their relation to fluxbranes. 

\subsection{Holography and SymTFT}

\paragraph{The holographic field theory.} The $AdS/CFT$ correspondence, i.e., the holographic principle, is a conjectural duality between an $AdS_{d+1}$ gravitational theory and a $d$-dimensional conformal field theory (CFT) reside on the $AdS_{d+1}$ boundary $\partial(AdS_{d+1})$ \cite{Maldacena:1997re,Gubser:1998bc,Witten:1998qj}.

In the context of superstring/M-theory, the $AdS_{d+1}/CFT_{d}$ correspondence  is realized in terms of the back-reaction of a stack of D-branes \cite{Maldacena:1997re,Gubser:1998bc,Witten:1998qj}. The precise gravitational theory in $AdS_{d+1}$, and consequently the dual CFT on the boundary, depends on the internal geometry $L_{n}$,
\begin{equation}
    AdS_{d+1}\,\times L_{n}\,,\quad \,\,\text{with}\quad \,\,D\,=\,d+1+n\,.
\end{equation}
Here, $D$ denotes the total spacetime dimension of the superstring/M-theory. We will also refer to the internal space $L_{n}$ as the link space. 

\begin{figure}[H]
\centering
\begin{tikzpicture}
\node[above right] (img) at (0,0) {\includegraphics[width=0.68\textwidth]{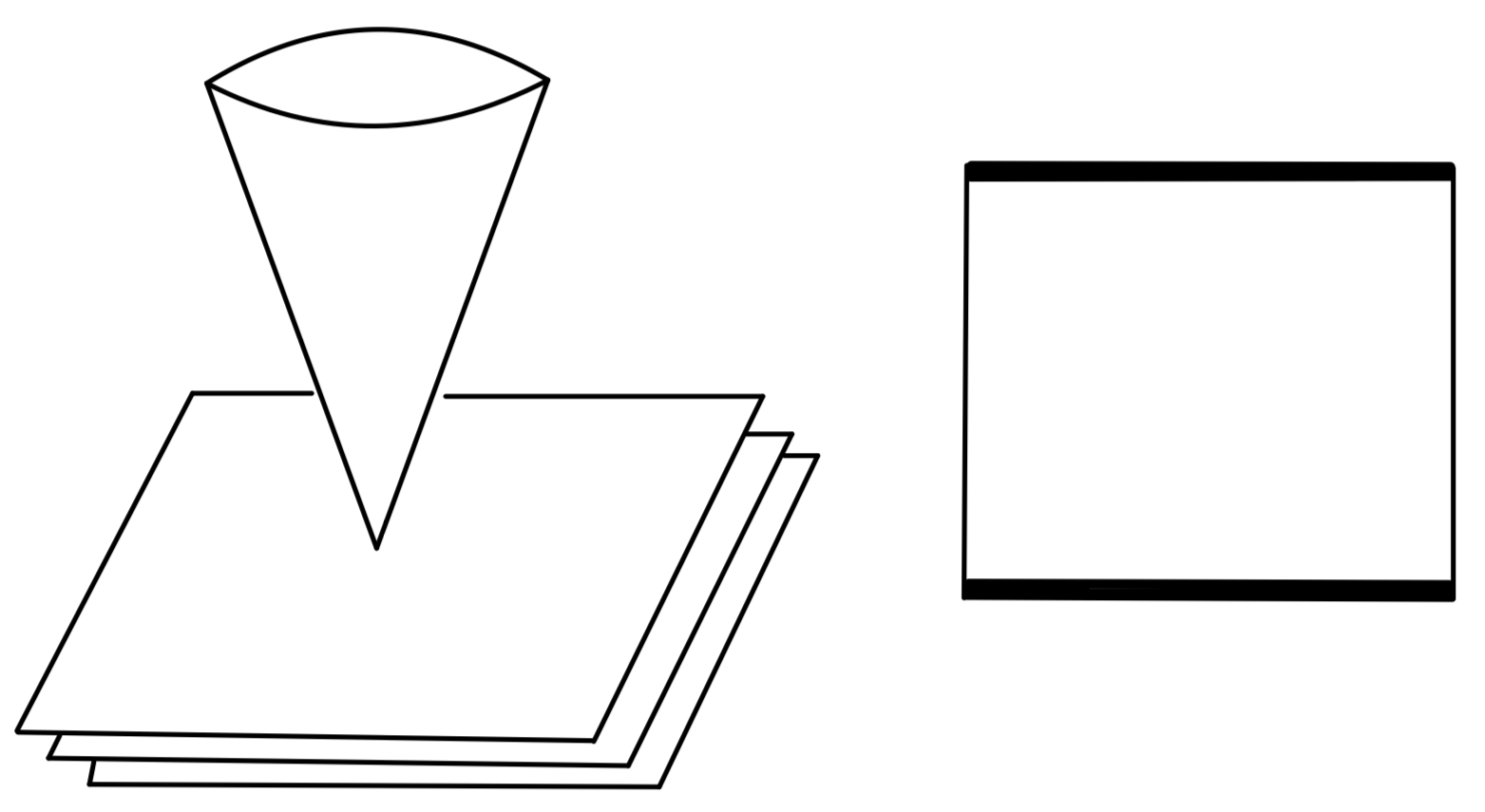}};
\node at (75pt,40pt){{stack of $N$ D-branes}};
\node at (240pt,30pt){{Physical boundary $\mathcal{B}^{\scriptscriptstyle\text{phys}}$}};
\node at (15pt,120pt){{$X=\text{cone}(L)$}};
\node at (245pt,90pt){{SymTFT}};
\node at (90pt,175pt) {Link space $L$};
\node at (245pt,150pt) {Symmetry boundary $\mathcal{B}^{\scriptscriptstyle\text{sym}}$};
\end{tikzpicture}
\caption{On the left, we have $N$ D-branes placed at the apex of the cone space $X_{n+1}$. On the right, we present the SymTFT theory along with its boundaries}
\label{Fig:AdS-CFT-Branes-SymTFT}
\end{figure}
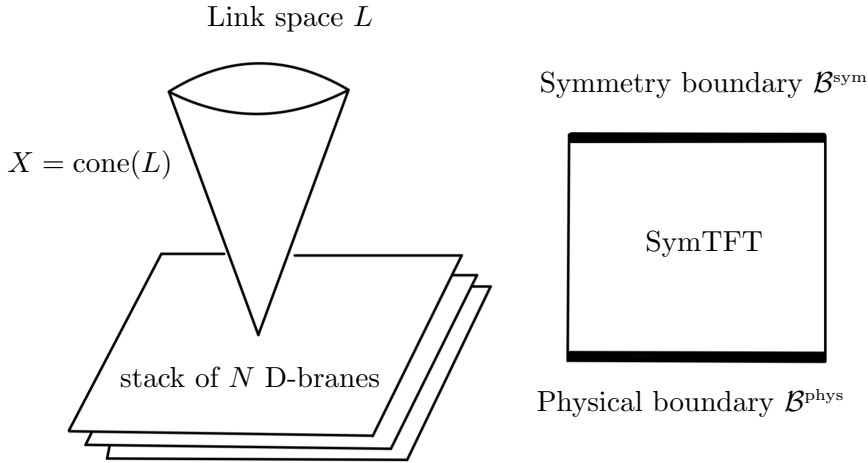

The field theories on the $AdS$ boundary, and generalizations even beyond conformal ones, can be seen to reside on D-branes occupy a $M_{d}=\R^{1,d-1}$ space and placed at the tip of conical geometry \cite{Klebanov:2000me,b22b0c4669d14657bc69fa68eb4776ea},
\begin{equation}
    X_{n+1}\,=\,\text{Cone}(L_{n})\,.
\end{equation}
To further break supersymmetry, one may take $X_{n+1}$ to admits special holonomy, e.g., \cite{Klebanov:1998hh,Klebanov:2000nc,Klebanov:2000hb,Strassler:2005qs}. The near-horizon limit of this realization forces the total geometry to take the form of $AdS_{d+1}\times L_{n}$ \cite{Maldacena:1997re,Gubser:1998bc,Witten:1998qj}. The physical picture of the above realization is given on the left-hand-side of Figure \ref{Fig:AdS-CFT-Branes-SymTFT}.

\paragraph{Setting the stage: SymTFT from holography.}

The conventional wisdom in obtaining bulk SymTFT theories within the framework of $AdS_{d+1}/CFT_{d}$ correspondence can be summarized by the following three key points:
\begin{itemize}
    \item Topological limit: The first step in deriving the SymTFT is to identify the topological sector of the higher-dimensional theory.  For generalized symmetries purposes, the focus is on the NS-NS and RR $p$-form field strengths \cite{Apruzzi:2021nmk,Apruzzi:2023uma}. Consequently, the gravitational aspects of the theory, as well as the dilaton field, can be neglected in this context.

    To implement this within the holographic framework, we follow the general approach outlined in \cite{Witten:1998wy}, which involves taking the limit near the $AdS$ boundary at infinity. In this limit, the theory can be truncated to its topological sector.
    \item Democratic formulation: The democratic formulation provides a framework in which NS-NS and RR field strengths, along with their corresponding brane sources, are treated on equal footing. This approach considers fluxes and their Hodge duals as independent field strengths. Consequently, both the Bianchi identities and equations of motion can be unified as Bianchi identities \cite{Cremmer:1998px,Marolf:2000cb}. This perspective enables the construction of an 11-dimensional topological theory (or a 12-dimensional extension in the context of M-theory) from which all Bianchi identities follow naturally \cite{Belov:2006xj,Apruzzi:2023uma}. Therefore, the topological theory, near the AdS boundary, treats all fields—electric and magnetic—equally.   
    \item Relation to SymTFT: As a final step, we reduce the higher-dimensional $(D+1)$-dimensional topological theory on the internal space $L_{n}$. The resulting is an auxiliary $(d+2)$ topological action, from which one can construct the bulk $(d+1)$-dimensional SymTFT \cite{Apruzzi:2023uma},
\begin{equation}\label{eq:from-top-to-aux-to-symtft}
    S_{d+2}^{\scriptscriptstyle{\text{Top}}}\,=\,\int_{ L_{n}}\,S^{\scriptscriptstyle\text{Top}}_{\scriptscriptstyle\text{(D+1)}} \qquad \longrightarrow\qquad S_{d+1}^{\scriptscriptstyle\text{SymTFT}}\,.
\end{equation}

With the realisation that the $d$-dimensional boundary theory lives on D-branes at the apex of the cone space $X_{n+1}$. Then the SymTFT can be effectively described to live along
\begin{equation}
    [0,\infty)\,\times\, M_{d}\,.
\end{equation}
Here, $[0,\infty)$ correspond to the radial direction of the cone $X_{n+1}$. The physical boundary $\mathcal{B}^{\scriptscriptstyle\text{phys}}$, of the bulk SymTFT, is located at $\{0\}\times M_{d}$. The above discussion is depicted in Figure \ref{Fig:AdS-CFT-Branes-SymTFT}. Therefore, the SymTFT acquires a similar realization to that given in the context of geometric engineering as, e.g., \cite{Apruzzi:2021nmk,Najjar:2024vmm}. 
\end{itemize}

Even though, in this work, we focus on non-conformal holographic field theories, as in \cite{Klebanov:2000hb}, the above strategy still applicable.  

Before closing this subsection, we review the type IIB supergravity action and its 11d topological limit. 

\paragraph{The type IIB supergravity.} To establish notation and provide a foundation for the subsequent discussion, let us briefly review the type IIB supergravity action. 

The bosonic part of the type IIB supergravity takes the following form \cite{Bergshoeff:1995as,Bergshoeff:1995sq,Cremmer:1998px,Polchinski:1998rr,Kaku:1999yd,ortin2004gravity,Becker:2006dvp,cecotti2023introduction}
\begin{equation}
\begin{split}
S_{\scriptscriptstyle\text{IIB}}\,\, = &\,\,\frac{1}{2\kappa^2} \int d^{10}x\, \sqrt{-g}\, e^{-2\Phi}\, \left[ R + 4 \partial_\mu \Phi \partial^\mu \Phi - \frac{1}{2} |\widetilde{H}_{3}|^2 \right]
\\
&- \frac{1}{4\kappa^2} \int d^{10}x\, \sqrt{-g}\, \left[ |\widetilde{F}_{1}|^2 + |\widetilde{F}_3|^2 + \frac{1}{2} |\widetilde{F}_5|^2 \right] - \frac{1}{4\kappa^2} \int C_4 \wedge H_{3} \wedge F_3\,.
\end{split}
\end{equation}
where $\kappa$ is the 10-dimensional gravitational coupling constant, $\Phi$ is the dilaton field, $H_{3}=dB_{2}$ is the NS-NS 3-form field strength, and $F_{p}=dC_{p-1}$ are the Ramond-Ramond (RR) $p$-form field strengths. The modified field strengths $\widetilde{F}_{p}$ are defined as
\begin{equation}\label{eq:def-flux-fields-IIB}
\widetilde{F}_{p} \,=\, dC_{p-1} \,-\, C_{p-3}\,\wedge\, H_{3}\,, \quad  p=1,3,5,7,9\,.
\end{equation}
Note that, $\widetilde{H}_{3}=H_{3}$ and $\widetilde{F}_{1}=F_{1}$. Equivalently, we can define $\widetilde{F}_{p}$ as \begin{equation}\label{eq:def-flux-fields-IIB(2)}
    \widetilde{F}_{p} \,=\, F_{p}\,+\, F_{p-2}\,\wedge\, B_{2}\,\quad p=1,3,5,7,9\,.
\end{equation}

The 5-form field strength $\widetilde{F}_{5}$ satisfies a self-duality condition. This condition is a supplement to the equation of motion that derived from the above action.

\paragraph{The topological 11-dimensional action.} The conventional approach to deriving the bulk SymTFT action involves utilizing the topological 11d type IIB action \cite{Apruzzi:2023uma}. To arrive at the 11d action, we ignore the gravitational part, and we set the dilaton $\Phi$ to zero. Furthermore, we use both electric and magnetic field strengths democratically \cite{Cremmer:1998px,Belov:2006xj,Bergshoeff:2001pv}. Therefore, we consider the Hodge stare operation on the field strengths, which are given as, e.g., \cite[Sec.3]{Apruzzi:2023uma},
\begin{equation}
    \widetilde{H}_{7} = \ast \widetilde{H}_{3}, \qquad \widetilde{F}_{p} = (-1)^{\lfloor{\frac{p}{2}}\rfloor}\, \ast \widetilde{F}_{10-p}\,.
\end{equation}
The on-shell Bianchi identities are then given as \cite{Cremmer:1998px}
\begin{equation}\label{eq:BI-Type-IIB}
\begin{split}
 dH_3 &= 0\,,
 \\
d\widetilde{F}_{p} + F_{p-2}\wedge H_{3} &= 0\,,
\\
d\widetilde{H}_{7} +  F_{1} \wedge \widetilde{F}_7 + F_{5} \wedge \widetilde{F}_3 &= 0\,.
\end{split}
\end{equation}
In fact, the second equation can be written as $d\widetilde{F}_{p} + \widetilde{F}_{p-2}\wedge H_{3} = 0$ as $H_{3}\wedge H_{3}=0$. The magnetic field strength $\widetilde{H}_{7}$ can be taken as \cite{ortin2004gravity},
\begin{equation}\label{eq:definition-H7-IIB}
    \widetilde{H}_{7} = H_{7} - C_{0}\wedge \widetilde{F}_{7} - C_{4}\wedge \widetilde{F}_{3} \,,\quad \text{with}\quad H_{7}= dB_{6}\,.
\end{equation}

The above Bianchi identities results from the following 11d type IIB topological action \cite{Belov:2006xj,Bergshoeff:2001pv,Apruzzi:2023uma,Tian:2024dgl}
\begin{equation}
\begin{split}
S_{\scriptscriptstyle\text{IIB}}^{\scriptscriptstyle\text{11d}} = &\int_{\mathcal{M}_{11}} \,\left[\widetilde{F}_{1}\wedge d\widetilde{F}_{9} - \widetilde{F}_{3}\wedge d\widetilde{F}_{7} +\frac{1}{2} \widetilde{F}_{5}\wedge d\widetilde{F}_{5} + H_{3}\wedge d\widetilde{H}_{7}\,\right]\, 
\\
\,+\,&\int_{\mathcal{M}_{11}}\,\left[\, H_{3}\wedge \widetilde{F}_{1}\wedge \widetilde{F}_{7}\, - \, H_{3}\wedge \widetilde{F}_{3} \wedge \widetilde{F}_{5}\,\right]\,.
\end{split}
\end{equation}
The first line are seen as the 11d topological lifts of the kinetic terms of the corresponding IIB gauge fields, while the second is the lift of the IIB Chern–Simons terms. In our conventions, the partition function of the type IIB theory is given by $\exp(2\pi i S_{\scriptscriptstyle\text{IIB}}^{\scriptscriptstyle\text{11d}})$.

Finally, we present the following notes: 
\begin{itemize}
    \item To get the above form of the 11d action, the following field redefinitions are taking into consideration \cite[App.A]{Apruzzi:2023uma},
\begin{equation}
    \widetilde{F}_{p}\ \to \ (2\pi l_{s})^{p-1} \,F_{p}\,,\qquad\, H_{3}\ \to \ (2\pi l_{s})^{2} H_{3}\,,\qquad \, \widetilde{H}_{7} \ \to \ (2\pi l_{2})^{-2}\, \widetilde{H}_{7}\,. 
\end{equation}
This is to ensure that we have integer periods for the various $p$-form flux fields. After performing the redefinitions, we set $l_{s} = 1$.

\item Using $H_{3}\wedge H_{3}=0$, one can show that the Chern–Simons terms satisfy,
\begin{equation}\label{eq:CS-term-1}
    H_{3}\wedge\widetilde{F}_{1}\wedge\widetilde{F}_{7} =  H_{3}\wedge{F}_{1}\wedge\widetilde{F}_{7}=  H_{3}\wedge\widetilde{F}_{1}\wedge{F}_{7} = H_{3}\wedge{F}_{1}\wedge{F}_{7} \,,
\end{equation}
and
\begin{equation}\label{eq:CS-term-2}
    H_{3}\wedge\widetilde{F}_{3}\wedge\widetilde{F}_{5} = H_{3}\wedge{F}_{3}\wedge\widetilde{F}_{5} =H_{3}\wedge\widetilde{F}_{3}\wedge{F}_{5} = H_{3}\wedge{F}_{3}\wedge{F}_{5}\,.
\end{equation}
\end{itemize} 

\subsection{Defects and symmetry operators}

Since the concept of generalized symmetries was embedded within the framework of superstring theory and M-theory, various works have explored the brane realization of charged defects and topological symmetry operators. Key contributions include \cite{Apruzzi:2021phx,Apruzzi:2022rei,Heckman:2022muc,Cvetic:2023plv,Apruzzi:2023uma,Garcia-Valdecasas:2023mis,Waddleton:2024iiv,Najjar:2024vmm,Najjar:2025rgt}, which have significantly advanced our understanding of these structures. In particular, the following discussion relies extensively on our previous work in \cite{Najjar:2024vmm, Najjar:2025rgt}.

For completeness, we provide a concise summary of the realization of defects and symmetry operators in the context of the $AdS_{d+1}/CFT_{d}$ correspondence. This discussion is guided by the physical picture illustrated in Figure \ref{Fig:AdS-CFT-Branes-SymTFT}.

\paragraph{Charged defects.}

On the holographic field theory which reside on the D-branes of Figure \ref{Fig:AdS-CFT-Branes-SymTFT}, i.e., on the boundary theory, charged objects under a given $m$-form symmetry, for both continuous or discrete symmetries, which are heavy non-dynamical $m$-dimensional objects. In general, these objects have realization in terms of (BPS) branes wrapping cycles of the link space $L_{n}$ and extend along the radial direction $\R_{+}$ of $X_{n+1}$ \cite{DelZotto:2015isa,Albertini:2020mdx}, 
\begin{equation}\label{def:defect1}
    \mathbb{D}^{m} := \bigcup_{p=-1,1,3,5,7} \{\text{BPS $p$-branes on } H_{p-m}(L_{9-d},\Z) \times \R_{+}\, \} \,.
\end{equation}
Equivalently, these branes are extended along the radial direction of the $AdS_{d+1}$ space. In the $AdS_{d+1}$ gravitational theory, these defects are lively stringy objects, see, e.g., \cite{Apruzzi:2021phx,Apruzzi:2022rei}.  

The $m$-dimensional defect operator is determined via the Wess-Zumino action of the BPS $p$-branes as
\begin{equation}
    \mathcal{O}(\Sigma_{m}) \,=\, \exp(in\, S_{\scriptscriptstyle\text{WZ}}^{\scriptscriptstyle(p+1)})\,\simeq\, \exp(in\,\int_{\Sigma_{m}\times \gamma_{p-m}\times \R_{+}} \,C_{p+1})\,,\,\quad n\in\Z\,.
\end{equation}

\paragraph{Symmetry topological operators.}

In general, $m$-dimensional defect of a $d$-dimensional theory is charged under $m$-form symmetry which is supported over co-dimension $(m+1)$, i.e., $(d-m-1)$-dimensional, topological operator \cite{Gaiotto:2014kfa}. These symmetry operators have brane realization which are given by $p$-branes, not necessary BPS, wrapping some cycles of the link space and extend transversely to the radial direction $\R_{+}$ \cite{Heckman:2022muc,Cvetic:2023plv}. Furthermore, a defect $\mathcal{O}(\Sigma_{m})$ is said to be charged under a symmetry topological operator $\mathcal{U}(\Sigma_{d-m-1})$ if they have non-trivial pairing both in spacetime and the link space $L_{n}$. The set of all such $(m'+1)$-dimensional topological operators is, 
\begin{equation}\label{def:symmetryoperator1}
    \mathbb{U}^{m'+1} =\bigcup_{p} \,\, \{\,\text{$p$-branes wrapping $H_{p-m'}(L_{9-d},\Z)$ and transverse to $\R_{+}$\,}\}\,.
\end{equation}
Here, $m'+1=d-m-1$.

In the context of $AdS/CFT$ construction, the brane realization of symmetry topological operators is determined by the brane-type. In particular, we have
\begin{itemize}
    \item Discrete $m$-form symmetries are realized by BPS branes, i.e. R-R D$p$-branes or NS-NS $p$-branes, wrapping free cycles of the link $L_{9-d}$ \cite{Apruzzi:2021phx,Apruzzi:2022rei}.
    \item Continuous $m$-form symmetries are realized via fluxbranes, or equivalently Page charges as we will emphasise soon, wrapping free cycles of the link $L_{9-d}$. 
\end{itemize}
Again, in the gravitational $AdS_{d+1}$ theory, these objects are stringy objects, e.g., F1-strings, D$p$-brane, fluxbranes, etc. 

We now turn our attention to the equivalence between fluxbranes and Page charges, at least for the cases of interest. 

\subsection{Fluxbranes, Page charges and symmetry operators}

\paragraph{Page charges.}

Following the general discussion in \cite{Page:1983mke,Bremer:1997qb,Marolf:2000cb}, \cite[App.A]{Chiodaroli:2010mv}, \cite[App.D]{deBoer:2012ma} and \cite[App.B]{Chowdhury:2013pqa}, the Page charges of the various strings and branes of the type IIB theory can be derived from the Bianchi identities in (\ref{eq:BI-Type-IIB}). In particular, we treat the Bianchi identities as conserved currents $d\ast J=0$, with $\ast J$ being the currents. To obtain the Page charges, we integrate the currents $J$ over a surface $\Sigma_{p}$. Schematically, the Page charge is defined as
\begin{equation}
    Q^{\bullet}_{p} := \int_{\Sigma_{p}}\,\ast J_{d-p}\,.
\end{equation}
Here, $\bullet$ indicates the brane's type (e.g., D-branes, NS5-branes, etc). The Page charge is both quantized and conserved, making it a suitable candidate for defining symmetry topological operators. We will refer to the integrand of the above Page charges as $P_{p}^{\bullet}$.

Therefore, it follows from (\ref{eq:BI-Type-IIB}) that the type IIB Page charges are given as
\begin{equation}\label{eq:def-Page-charges}
    \begin{split}
        Q_{3}^{\scriptscriptstyle\text{NS5}} &\,=\, \int_{\Sigma_{3}}\,H_{3}\,,
        \\
        Q_{p}^{\scriptscriptstyle\text{D$(8-p)$}} &\,=\, \int_{\Sigma_{p}}\,\left( \widetilde{F}_{p}+ C_{p-3}\wedge H_{3}\right)\,\,,
        \\
        Q_{7}^{\scriptscriptstyle\text{F1}} &\,=\, \int_{\Sigma_{7}}\,\left(\widetilde{H}_{7} + C_{0}\wedge \widetilde{F}_{7} + C_{4}\wedge \widetilde{F}_{3}\right)\, .
    \end{split}
\end{equation}

\paragraph{Page charges as fluxbranes.}

Consider the page charge associated with the D1-brane, where the integrand is given as
\begin{equation}
  P_{7}^{\scriptscriptstyle\text{D1}} \,=\,  \widetilde{F}_{7} + C_{4}\wedge H_{3}\,.
\end{equation}
One can show that $d\widetilde{F}_{7}$ satisfies
\begin{equation}
    \begin{split}
        d\widetilde{F}_{7} \,&=\, - \widetilde{F}_{5}\wedge H_{3}
        \\
        \,&=\, d (\widetilde{F}_{5}\wedge B_{2} -\frac{1}{2}\widetilde{F}_{3}\wedge B_{2}\wedge B_{2} + \frac{1}{6}\widetilde{F}_{1}\wedge B_{2}\wedge B_{2}\wedge B_{2})\,.
    \end{split}
\end{equation}
Therefore, the integrand of the Page charge can be rewritten as
\begin{equation}
    P_{7}^{\scriptscriptstyle\text{D1}} \,=\,\widetilde{F}_{7} - \widetilde{F}_{5}\wedge B_{2} +\frac{1}{2} \widetilde{F}_{3}\wedge B_{2}\wedge B_{2} - \frac{1}{6}  \widetilde{F}_{1}\wedge B_{2}\wedge B_{2}\wedge B_{2}\,. 
\end{equation}
Hence, the $Q_{7}^{\text{D1}}$ Page charge is given by
\begin{equation}
   Q_{7}^{\text{D1}} = \int_{\Sigma_{7}} \,\sum_{i=0,2,4,6}\,\widetilde{F}_{7-i}\,\wedge\,e^{-B_{2}}\,.
\end{equation}
Similar expressions to the above Page charge can be found in \cite[(2.26)]{Benini:2007gx} and \cite[App.B]{Chowdhury:2013pqa}.

For a generic D$(8-p)$-brane, the Page charge can be written as
\begin{equation}\label{eq:Qp-D-brane-charge(2)}
    Q_{p}^{\scriptscriptstyle\text{D$(8-p)$}}\,=\int_{\Sigma_{p}}\, \sum_{i}\, \widetilde{F}_{p-i}\, \wedge\, e^{-B_{2}} \,.
\end{equation}
Putting the Page charge of the usual D-brane in this form, then it coincides with the (extended) topological Wess-Zumino Lagrangian as
\begin{equation}\label{eq:topological-action-Dp-2-brane}
\begin{split}
S^{\scriptscriptstyle\text{D$(p-2)$}}_{\scriptscriptstyle\text{Top}} &\,=\, \int_{\Sigma_{p}} d\widetilde{\mathcal{L}}_{\,\scriptscriptstyle\text{WZ}}^{\,\scriptscriptstyle\text{D$(p-2)$}}
\\
&\,=\, \int_{\Sigma_{p}} \, \sum_{i=0}^{p} \, \widetilde{F}_{p-i} \wedge e^{F_{2}-B_{2}}
\end{split}
\end{equation}
Here, $F_{2}$ is the field strength of the $U(1)$ gauge field on the world volume of the D-brane. The above equation matches (\ref{eq:Qp-D-brane-charge(2)}) when setting $F_{2}=0$, i.e.,
\begin{equation}
    \restr{S^{\scriptscriptstyle\text{D$(p-2)$}}_{\scriptscriptstyle\text{Top}}}{F_{2}=0} \,=\, Q_{p}^{\scriptscriptstyle\text{D$(8-p)$}}\,.
\end{equation}
Here, the condition $F_{2}=da_{1}=0$ ensures that we are dealing with invertible symmetry.

With the above identification, then it is clear that the fluxbranes defined in the sense of \cite{Cvetic:2023plv} can be interpreted as Page charges. In particular, one way to define the fluxbrane is to consider \cite[(2.10)]{Cvetic:2023plv},
\begin{equation}
    \int_{X_{10}}\,\widetilde{\mathcal{L}}_{\scriptscriptstyle\text{WZ}}^{\,\scriptscriptstyle\text{D$(p-2)$}}\,\wedge \, d\ast \widetilde{F}_{p} \,\sim\,  (-1)^{p}\,\int_{X_{10}} d\widetilde{\mathcal{L}}_{\,\scriptscriptstyle\text{WZ}}^{\,\scriptscriptstyle\text{D$(p-2)$}}\,\wedge\, \ast \widetilde{F}_{p}\,.
\end{equation}
According to \cite{Gutperle:2001mb}, the transverse flux $\ast \widetilde{F}_{p}$ should satisfy
\begin{equation}
    \ast \widetilde{F}_{p} \,=\, \alpha\,\mathsf{PD}{(\Sigma_{p})}\,.
\end{equation}
The parameter $\alpha$ can be seen to reside on the worldvolume of the fluxbrane. Therefore, the (invertible) symmetry topological operator, arise from D-branes, can be taken as
\begin{equation}
\begin{split}
        \mathcal{U}_{\alpha}^{\,\scriptscriptstyle\text{D$(8-p)$}}(\Sigma_{p})\,&=\, \exp(2\pi i \alpha\, Q_{p}^{\,\scriptscriptstyle\text{D$(8-p)$}})\,
        \\
        &=\,\exp(2\pi i \alpha\,\int_{\Sigma_{p}} P_{p}^{\,\scriptscriptstyle\text{D$(8-p)$}})\,
\end{split}
\end{equation}
Since $Q_{p}^{\bullet}\in \Z$, then $\alpha$ take values in $[0,1)$.

A similar perspective on the equivalence between the topological action of fluxbranes and Page charges is presented in \cite{Najjar:2024vmm}. This work, which focuses on M-theory and geometric engineering, demonstrates that by adopting the appropriate 7-dimensional fluxbrane topological action—subject to the conditions that the integrand is both quantized and closed—one arrives at the Page charge of the M2-brane. This suggests that the correspondence between the fluxbrane topological action and Page charges may extend beyond D-branes to other types of branes, including F1-strings, NS5-branes, KK-monopoles, and even exotic branes. Similar findings have been noted in recent studies, as discussed in \cite{Cvetic:2025kdn}.

Building on this insight, we next demonstrate that the appropriate topological action of the $H_{7}$-fluxbrane corresponds to the Page charge of the F1-string.

\paragraph{NS-NS fluxbrane.} Let us focus on the NS-NS fluxbrane and demonstrate the equivalence of its extended topological action to the F1-string Page charge defined in (\ref{eq:def-Page-charges}). The existence of NS-NS fluxbranes has been extensively discussed in the literature, particularly in \cite{Russo:1998xv,Costa:2000nw} and references therein.

The Wess-Zumino (WZ) action for the solitonic NS5B-brane can be expressed as \cite{Eyras:1998hn,Bergshoeff:2011zk,deBoer:2012ma}
\begin{equation}
    \begin{split}
S_{\scriptscriptstyle\text{NS5B}}^{\scriptscriptstyle\text{WZ}} &\,\sim\,  \int_{M_{6}}\, \left[\, B_6 \,+\, \frac{1}{2}\,(\,-G_6 \wedge C_0 + G_4 \wedge C_2 - G_2 \wedge C_4 + G_0 \wedge C_6\,)\, \right] \,,
 \end{split}
\end{equation}
where $M_{6}$ is the worldvolume of the NS5B-brane and $B_{6}$ is the dual gauge field of the NS-NS $B_{2}$ 2-form. The field strengths $G_{\bullet}$ are defined through the pull-back of the type IIB RR gauge fields and the worldvolume fields as
\begin{equation}
    G_0 = C_0, \quad G_2 = C_2 + dc_1, \quad G_4 = C_4 + dc_3 - H_3 \wedge c_1, \quad G_6 = C_6 + dc_5 - H_3 \wedge c_3.
\end{equation}
Here, $c_{1}, c_{3}, c_{5}$ are worldvolume gauge fields. For further details on the worldvolume theory of the NS5B-brane, see \cite{Callan:1991ky}.

Since we are interested in constructing invertible symmetry operators, it is convenient to turn off the worldvolume gauge fields. In this limit, the topological action simplifies to
\begin{equation}
   S_{\scriptscriptstyle\text{NS5B}}^{\scriptscriptstyle\text{WZ}} \,\sim\,  \int_{M_{6}}\,  B_6 \,. 
\end{equation}
This action can be extended to a seven-dimensional description of the $H_{7}$-fluxbrane,
\begin{equation}
   S_{\scriptscriptstyle\text{NS5B}}^{\scriptscriptstyle\text{Top}} \,\sim\,  \int_{\Sigma_{7}}\,  H_{7} \,,
\end{equation}
where $H_{7}=dB_{6}$. Using (\ref{eq:definition-H7-IIB}), the fluxbrane action can be rewritten as 
\begin{equation}
   S_{\scriptscriptstyle\text{NS5B}}^{\scriptscriptstyle\text{Top}} \,\sim\,   \int_{\Sigma_{7}}\,\left(\widetilde{H}_{7} + C_{0}\wedge \widetilde{F}_{7} + C_{4}\wedge \widetilde{F}_{3}\right)\, ,
\end{equation}
which coincide with $Q_{7}^{\scriptscriptstyle\text{F1}}$ of (\ref{eq:def-Page-charges}). In other words, we conclude that
\begin{equation}
\restr{S_{\scriptscriptstyle\text{NS5B}}^{\scriptscriptstyle\text{Top}}}{\forall c_{i}=0}\,=\,Q_{7}^{\scriptscriptstyle\text{F1}}\,.
\end{equation}

\section{\texorpdfstring{$(-1)$}{(-1)}-form symmetries in the KS set-up}\label{sec:(-1)-form-KS-theory}

The Klebanov-Strassler (KS) solution offers a holographic realization of pure 4d $\N=1$ $SU(M)$ Super Yang-Mills (SYM) theory. In this section, we aim to systematically investigate the possible continuous $(-1)$-form symmetries of the gauge theory. We will derive the corresponding SymTFT and demonstrate how one can consistently define a modified instanton sum. Furthermore, we will show how a 4-group structure, i.e., decomposition, naturally emerges within this framework.

\subsection{The KS set-up, a lightning review}

Before delving into the $(-1)$-form symmetry of the KS solution \cite{Klebanov:2000hb} and its physical implications, we first provide a brief review of the theory, highlighting its key features. The KS set-up is formulated within type IIB superstring theory, building on earlier works \cite{Klebanov:1998hh, Klebanov:2000nc}.

With the physical picture of Figure \ref{Fig:AdS-CFT-Branes-SymTFT} in mind, the holographic dual field theory is described in terms of D3- and D5-branes as follows:
\begin{itemize}
    \item Place $N$ D3-branes at the tip of the CY3 conifold \cite{Candelas:1989js} and extend along the 4d spacetime $\R^{1,3}$. The D3-branes source the self-dual $\widetilde{F}_{5}$ flux, and the presence of the D3-branes can be felt by measuring the 5-form flux at the link space $T^{1,1}$,
    \begin{equation}\label{eq:KS-F5-flux}
        \int_{T^{1,1}}\,\widetilde{F}_{5}\,=\,N\,.
    \end{equation}
    The near-horizon limit of the above configuration is given as $AdS_{5}\times T^{1,1}$.
    \item Furthermore, there are $M$ fractional 3-branes which originate from $M$ D5-branes wrapping $S^{2}\subset T^{1,1}$. The presence of the D5-branes can be measured at the link space by considering
    \begin{equation}
        \int_{S^{3}}\,\widetilde{F}_{3}\,=\,M\,.
    \end{equation}
    The existence of the D5-branes modifies the $AdS_{5}\times T^{1,1}$ background. In particular, they modify the metric on the $AdS_{5}$ part, and we denote the gravitational bulk spacetime by ${M}_{5}$. The details of the modified metric can be found in \cite{Klebanov:2000hb}; however, as they are not relevant to the present work, we do not include them in this review. Furthermore, the 5-form flux is modified as
    \begin{equation}
    \int_{T^{1,1}}\,\widetilde{F}_{5}\,=\,N\,+\, \,\frac{3g_{s}M^{2}}{2\pi}\,\ln(r/r_{0})\,,
    \end{equation}
    which is only quantized at special values of the radial distance $r$. Here, $r_{0}$ can be interpreted as a UV scale of the field theory \cite{Klebanov:2000hb}.
\end{itemize}

The holographic field theory description is given by a 4d $\N=1$ $SU(N+M)\times SU(N)$ gauge theory with bifundamental matter fields transforming as $(N+M,\overline{N})$ and $(\overline{N}+\overline{M},N)$. To leading order, the gauge couplings of the two gauge group factors run in opposite directions \cite{Klebanov:2000hb,Strassler:2005qs},
\begin{equation}
    \frac{\partial}{\partial\log(\mu)}\frac{8\pi^{2}}{g_{1}^{2}}\,\sim\, 3M+\cdots\,,\qquad \frac{\partial}{\partial\log(\mu)}\frac{8\pi^{2}}{g_{2}^{2}}\,\sim\, -3M+\cdots\,.
\end{equation}
 In particular, the $SU(N+M)$ gauge group becomes strongly coupled, causing $g_{1}$ to diverge, while $g_{2}$ of the $SU(N)$ gauge group decreases. Since the $SU(N+M)$ gauge factor has $2N$ flavours, one can continue past this regime by applying Seiberg duality \cite{Seiberg:1994pq}, which shifts $N+M$ as
\begin{equation}
    N+M\ \to \ N-M\,.
\end{equation}
This duality generates massive mesons that can be integrated out. Iterating this process leads to the following duality cascade \cite{Klebanov:2000hb,Strassler:2005qs},
\begin{equation}
   SU(N+M) \times SU(N) \to  SU(N) \times SU(N-M) \to SU(N-M) \times SU(N-2M) \cdots\,.
\end{equation}
After $k$ steps, the gauge group takes the form
\begin{equation}
    SU(N+M-kM)\,\times\,SU(N-kM)\,,\qquad  \text{(after $k$ steps)}\,.
\end{equation}
If the number of D3-branes is an integer multiple of the number of fractional 3-brane, i.e.,
\begin{equation}
    N\,=\,\kappa\,M\,,
\end{equation}
then after $\kappa$ steps, the theory flows to a pure gauge theory description,
\begin{equation}
   \text{Pure\, 4d\, $\N=1$\, $SU(M)$\, SYM} \qquad  \text{(after $\kappa$ steps)}\,.
\end{equation}

Before analysing how the above pure gauge theory decomposes, we first determine all possible $(-1)$-form symmetries of the theory.

\subsection{\texorpdfstring{$(-1)$}{(-1)}-form symmetries in the KS set-up}

In this subsection, we define the symmetry operators that generate continuous $(-1)$-form symmetries, analyse the relevant aspects of the bulk 5d SymTFT, and identify the $\theta_{\text{YM}}$ parameter of the gauge theory.

\paragraph{$(-1)$-form symmetry operators.}

Symmetry topological operators for $(-1)$-form symmetries, whether discrete or continuous, are defined over the entire spacetime manifold $M_{d}$ \cite{Vandermeulen:2022edk,Aloni:2024jpb,Santilli:2024dyz}. In the context of string theory, the realization of such operators for continuous symmetries—whether in the $AdS/CFT$ framework or through geometric engineering—requires the presence of at least one $(d+r)$-dimensional fluxbrane wrapping an $r$-cycle of the link space $L_{n}$ \cite{Najjar:2024vmm}. 

Therefore, to determine the continuous $(-1)$-form symmetry operators of the KS solution, we should first consider the (co)homology groups of the link $T^{1,1}$ geometry,
\begin{equation}
        H_{\bullet}(T^{1,1})\,=\, H^{5-\bullet}(T^{1,1})\,=\,(\,\Z\,, 0\,,\Z\,,\Z\,,0,\Z\,)\,.
\end{equation}
It is clear that one can construct three different symmetry topological operators for the continuous $(-1)$-form symmetry, which are given as:
\begin{itemize}
    \item Wrapping the Page charge of the D$(-1)$-brane, i.e., the F9-fluxbrane (or $P_{9}^{\scriptscriptstyle\text{D(-1)}}$), on the fifth homological cycle which correspond to the whole link space $T^{1,1}$. In this case, the charged defect operator would morally be given by the D$(-1)$-brane which extend along the radial direction and occupy $(-1)$-direction in the 4d spacetime. Of course, $(-1)$-form symmetries have no charged defects \cite{Aloni:2024jpb,Santilli:2024dyz,Najjar:2024vmm}; however, it is convenient to point out their stringy origin. The D$(-1)$-brane appears as a $(-1)$-brane in the gravitational bulk ${M}_{5}$ spacetime. 

The symmetry topological operator is given as:
\begin{equation}\label{eq:sym-top-op-P9}
\begin{split}
     \mathcal{U}^{\scriptscriptstyle \text{D(-1)}}_{\alpha_{\scriptscriptstyle(1)}}(\Sigma_{4})  \,&=\,   \exp{i\frac{\alpha_{\scriptscriptstyle(1)}}{2\pi}\,\int_{\Sigma_{4}\times T^{1,1}}\,P_{9}^{\scriptscriptstyle\text{D(-1)}}}
     \\
     \,&\equiv\,  \exp{i\frac{\alpha_{\scriptscriptstyle(1)}}{2\pi}\,\int_{\Sigma_{4}}\,F_{4}^{(9)}}\,.
\end{split}
\end{equation}
The precise form of $F_{4}^{(9)}$ would be given by dimensionally reducing $P_{9}^{\scriptscriptstyle\text{D(-1)}}$ in terms of the cohomological bases of the internal $T^{1,1}$ space. Its form would be given in (\ref{eq:def-F49}). 
    \item Wrapping the Page charge of the D1-brane, i.e., the F7-fluxbrane, on the third homological cycle $S^{3}\subset T^{1,1}$. In the ${M}_{5}$ spacetime, the charged defect appears as another $(-1)$-brane, which originate from D1-brane wrapping $S^{2}\subset T^{1,1}$. Applying the definition of the charged defects (\ref{def:defect1}), we recover the corresponding 4d spacetime charged operator.

    The corresponding symmetry operator is given as:
    \begin{equation}
\begin{split}
     \mathcal{U}^{\scriptscriptstyle \text{D1}}_{\alpha_{\scriptscriptstyle(2)}}(\Sigma_{4})  \,&=\,   \exp{i\frac{\alpha_{\scriptscriptstyle(2)}}{2\pi}\,\int_{\Sigma_{4}\times S^{3}}\,P_{7}^{\scriptscriptstyle\text{D1}} }\,
     \\
     & =\, \exp{i\frac{\alpha_{\scriptscriptstyle(2)}}{2\pi}\,\int_{\Sigma_{4}\times T^{1,1}}\, \mathsf{PD}{(S^{3})}\wedge P_{7}^{\scriptscriptstyle\text{D1}}}
     \\
     &\equiv\,  \exp{i\frac{\alpha_{\scriptscriptstyle(2)}}{2\pi}\,\int_{\Sigma_{4}}\,F_{4}^{(7)}}\,.
\end{split}
\end{equation}
Here, $\mathsf{PD}{(S^{3})}$ is the Poincaré dual of the 3-cycle $S^{3}\subset T^{1,1}$. We define $F_{4}^{(7)}$ through (\ref{eq:def-F47}). 
    
    \item The third possibility is given by wrapping the Page charge of the F1-brane, i.e., the H7-fluxbrane, on the third homological cycle $S^{3}\subset T^{1,1}$. In the ${M}_{5}$ spacetime, the charged defect appears as another $(-1)$-brane, which originate from F1-brane wrapping $S^{2}\subset T^{1,1}$. Again, this would lead to the boundary 4d spacetime charged defect.

    In this case, the symmetry operator is written as:
    \begin{equation}\label{eq:sym-top-op-H7}
\begin{split}
     \mathcal{U}^{\scriptscriptstyle \text{F1}}_{\alpha_{\scriptscriptstyle(3)}}(\Sigma_{4})  \,&=\,   \exp{i\frac{\alpha_{\scriptscriptstyle(3)}}{2\pi}\,\int_{\Sigma_{4}\times S^{3}}\,P_{7}^{\scriptscriptstyle \text{F1}}}
      \\
     & =\, \exp{i\frac{\alpha_{\scriptscriptstyle(3)}}{2\pi}\,\int_{\Sigma_{4}\times T^{1,1}}\, \mathsf{PD}{(S^{3})}\wedge P_{7}^{\scriptscriptstyle \text{F1}}}
     \\
     &\equiv\,  \exp{i\frac{\alpha_{\scriptscriptstyle(3)}}{2\pi}\,\int_{\Sigma_{4}}\,H_{4}^{(7)}}\,.
\end{split}
\end{equation}
The form of the current $H_{4}^{(7)}$ is given by (\ref{eq:def-H47}).  
\end{itemize}

\paragraph{The currents structure.} 

Let us now determine the structure of the currents $F_{4}^{(9)}$, $F_{4}^{(7)}$, and $H_{4}^{(7)}$ that generate the corresponding $(-1)$-form symmetries. First, we expand the type IIB fluxes in terms of the cohomological bases of the internal $T^{1,1}$ geometry, which can be taken as
\begin{equation}
    H^{\bullet}(T^{1,1})\,=\,\{1\}\,,\{\omega_{2}\}\,,\{\omega_{3}\}\,,\{\omega_{5}\}\,.
\end{equation}
Here, we have $\omega_{2}\wedge\omega_{3}=\omega_{5}$, which reflects the topology of the $T^{1,1}=S^{2}\times S^{3}$.

The field strength expansion along these bases can be written as
\begin{equation}\label{eq:fluxes-expansion}
    \begin{split}
        F_{1} &= f_{1}^{(1)}\wedge 1\,,
        \\
        H_{3} &= h_{3}^{(3)}\wedge 1 + h_{1}^{(3)}\wedge\omega_{2}+h_{0}^{(3)}\wedge\omega_{3}\,,
        \\
        F_{3} &= f_{1}^{(3)}\wedge 1 + f_{1}^{(3)}\wedge\omega_{2}+f_{0}^{(3)}\wedge\omega_{3}\,,
        \\
        F_{5} &= f_{5}^{(5)}\wedge 1 + f_{3}^{(5)}\wedge\omega_{2}+f_{2}^{(5)}\wedge\omega_{3} + f_{0}^{(5)}\wedge\omega_{5}\,,
        \\
         H_{7} &= h_{7}^{(7)}\wedge 1 + h_{5}^{(7)}\wedge\omega_{2}+h_{4}^{(7)}\wedge\omega_{3} + h_{2}^{(7)}\wedge\omega_{5}\,,
         \\
          F_{7} &= f_{7}^{(7)}\wedge 1 + f_{5}^{(7)}\wedge\omega_{2}+f_{4}^{(7)}\wedge\omega_{3} + f_{2}^{(7)}\wedge\omega_{5}\,,
          \\
           F_{9} &= f_{9}^{(9)}\wedge 1 + f_{7}^{(9)}\wedge\omega_{2}+f_{6}^{(9)}\wedge\omega_{3} + f_{4}^{(9)}\wedge\omega_{5}\,.
    \end{split}
\end{equation}
The above expansion works for the closed $F_{p}$ and the non-closed $\widetilde{F}_{p}$ field strengths. We will add the tilde sign whenever it is needed.

In addition, we consider the expansion of the NS-NS 2-form gauge field $B_{2}$ as:
\begin{equation}
    B_{2} = b_{2}^{(2)}\,\wedge\, 1 \,+\, b_{0}^{(2)}\,\wedge\,\omega_{2}\,.
\end{equation}
Since $H_{3}=dB_{2}$, then it follows that:
\begin{equation}
    h_{3}^{(3)}\,=\,db_{2}^{(2)}\,,\qquad h_{1}^{(3)}\,=\,db_{0}^{(2)}\,.
\end{equation}
Furthermore, the reduction of the condition $H_{3}\wedge H_{3}=0$ implies
\begin{equation}
    h_{3}^{(3)}\wedge h_{3}^{(3)} =0\,.
\end{equation}

Apply the flux expansion in (\ref{eq:fluxes-expansion}) to the form of the Page charges as given in (\ref{eq:def-Page-charges}), then we can determine the field strengths $F_{4}^{(9)}$, $F_{4}^{(7)}$, and $H_{4}^{(7)}$. We will expand $dP^{\bullet}_{p}$ as it is more convenient. In particular, we first consider
\begin{equation}\label{eq:def-F49}
    \begin{split}
        dP_{9}^{\scriptscriptstyle\text{D(-1)}}\,&=\, d\widetilde{F}_{9} + F_{7}\wedge H_{3}\,
        \\
        \,&\quad \supset\, \left[d\widetilde{f}_{4}^{\,(9)} - f_{2}^{(7)}\wedge h_{3}^{(3)} - f_{4}^{(7)}\,\wedge\,h_{1}^{(3)} + f_{5}^{(7)}\,\wedge\,h_{0}^{(3)}\right]\,\wedge\,\omega_{5}\,
        \\
        &\qquad \equiv\, dF_{4}^{(9)}\,\wedge\,\omega_{5}\,.
    \end{split}
\end{equation}
The field strength $F_{4}^{(7)}$ is determined through
\begin{equation}\label{eq:def-F47}
    \begin{split}
        dP_{7}^{\scriptscriptstyle\text{D1}}\,&=\, d\widetilde{F}_{7} + F_{5}\wedge H_{3}
        \\
        \,&\quad\supset\, \left[ \, d\widetilde{f}_{4}^{\,(7)} + (f_{5}^{(5)}\wedge h_{0}^{(3)} - f_{2}^{(5)} \wedge h_{3}^{(3)}) \, \right]\wedge \omega_{3}\,
        \\
        &\qquad \equiv\, dF_{4}^{(7)}\,\wedge\,\omega_{3}\,.
    \end{split}
\end{equation}
Finally, we have
\begin{equation}\label{eq:def-H47}
    \begin{split}
        dP_{7}^{\scriptscriptstyle\text{F1}}\,&=\, d\widetilde{H}_{7} + F_{1}\wedge \widetilde{F}_{7} + F_{5}\wedge \widetilde{F}_{3}
        \\
        \,&\quad \supset\, \left[ \, d\widetilde{h}_{4}^{(7)} + f_{1}^{(1)}\wedge \widetilde{f}_{4}^{\,(7)} + (f_{5}^{(5)}\wedge \widetilde{f}_{0}^{\,(3)} - f_{2}^{(5)} \wedge \widetilde{f}_{3}^{\,(3)}) \, \right]\wedge \omega_{3}\,
        \\
        &\qquad \equiv dH_{4}^{(7)}\,\wedge\,\omega_{3}\,.
    \end{split}
\end{equation}

For completeness, let us also work out the reduction of the dual Page charges that correspond to NS5-brane, D5-brane, and D9-brane. Again, we expand the exterior derivative of the corresponding Page charges. For the NS5-brane, we have:
\begin{equation}\label{eq:def-H13}
        dP_{3}^{\scriptscriptstyle\text{NS5}}\,\, \supset dh_{1}^{(3)}\,\wedge\,\omega_{2}\, \equiv\, dH_{1}^{(3)} \qquad \Rightarrow\qquad   H_{1}^{(3)}\,=\,h_{1}^{(3)}\,. 
\end{equation}
In addition, for the D5-brane we write
\begin{equation}\label{eq:def-F13}
\begin{split}
            dP_{3}^{\scriptscriptstyle\text{D5}}\,\,&=\,d\widetilde{F}_{3} + F_{1}\,\wedge\,H_{3}\, 
            \\
            \,&\,\supset \left(\,d\widetilde{f}_{1}^{\,(3)} + f_{1}^{(1)}\wedge h_{1}^{(3)}\,\right) \,\wedge\,\omega_{2} \ \equiv\ dF_{1}^{(3)}\,\wedge\,\omega_{2}\, 
            \\
            &\Rightarrow \quad  F_{1}^{(3)}\,=\,\widetilde{f}_{1}^{\,(3)} + c_{0}^{(0)}\wedge h_{1}^{(3)} \,=\,f_{1}^{(3)}\,.
\end{split}
\end{equation}
To obtain the last equality we used a reduced version of (\ref{eq:def-flux-fields-IIB}). Finally, for the D9-brane, we have
\begin{equation} \label{eq:def-F11}
dP_{1}^{\scriptscriptstyle\text{D9}}\,\,\supset df_{1}^{(1)} \, \wedge \, 1 \, \equiv\, dF_{1}^{(1)}\,\wedge\,1\, \qquad \Rightarrow\qquad  F_{1}^{(1)} = f_{1}^{(1)}\,.
\end{equation}
The above Page charges can be used to construct three different continuous $2$-form symmetry topological operators.

Collectively, we refer to the lower-dimensional fluxes of the form $F_{\bullet}^{(\bullet)}$ or $H_{\bullet}^{(\bullet)}$ as lower-dimensional Page charges.

\paragraph{The continuous BF terms.}

Our approach to identifying the continuous BF terms is inspired by our work in \cite{Najjar:2024vmm}, where we construct the appropriate continuous BF terms using contributions from the topological CS terms after dimensional reduction.

Following the general discussion around (\ref{eq:from-top-to-aux-to-symtft}), we now construct the auxiliary 6d topological theory by integrating the 11d type IIB topological action on the link space $T^{1,1}$, 
\begin{equation}
    S_{\text{6d}}^{\scriptscriptstyle\text{Top}}\,=\,\int_{T^{1,1}}\,S_{\scriptscriptstyle\text{IIB-Top}}^{\scriptscriptstyle\text{11d}}\,.
\end{equation}

It is straight forward to show that the relevant "kinetic" terms reduced to 
\begin{equation}
   S_{\text{6d}}^{\scriptscriptstyle\text{Top}}\,\supset\, \int_{\mathcal{M}_{6}}\,f_{1}^{(1)}\,\wedge\,d\widetilde{f}_{4}^{\,(9)}\,-\,\widetilde{f}_{4}^{\,(7)}\,\wedge\,d\widetilde{f}_{1}^{\,(3)}\,+\, h_{1}^{(3)}\,\wedge\,d\widetilde{h}_{4}^{\,(7)}\,.
\end{equation}
On the other hand, the topological CS-terms would be reduced to
\begin{equation}
  \begin{split}
       H_{3}\wedge F_{1}\wedge F_{7} \,&=\, \left(h_{3}^{(3)}\wedge f_{1}^{(1)} \wedge f_{2}^{(7)}\right)\wedge \omega_{5}
        \\
        &\,+ \left(h_{1}^{(3)}\wedge f_{1}^{(1)} \wedge \widetilde{f}_{4}^{\,(7)}\right)\wedge\omega_{2}\wedge\omega_{3} 
        \\
        &\,+ \left(h_{0}^{(3)}\wedge f_{1}^{(1)} \wedge f_{5}^{(7)}\right)\wedge\omega_{3}\wedge\omega_{2}\,,
  \end{split}
\end{equation}
and
\begin{equation}
    \begin{split}
        H_{3}\wedge F_{3}\wedge F_{5} \,&\supset\, \left(h_{3}^{(3)}\wedge {f}_{1}^{(3)}\wedge f_{2}^{(5)} \,\,+\,\, h_{0}^{(3)}\wedge {f}_{1}^{(3)}\wedge f_{5}^{(5)}\right)\wedge\omega_{2}\wedge\omega_{3} 
        \\ 
        &\,\,\,+\left(h_{1}^{(3)}\wedge \widetilde{f}_{3}^{\,(3)}\wedge f_{2}^{(5)}\,\,-\,\,h_{1}^{(3)}\wedge \widetilde{f}_{0}^{\,(3)}\wedge f_{5}^{(5)}\right)\wedge\omega_{2}\wedge\omega_{3} \,.
    \end{split}
\end{equation}
Here, we have used the properties of the CS-terms given in (\ref{eq:CS-term-1}) and (\ref{eq:CS-term-2}). 

Combine the above results, the auxiliary 6d topological theory of interest is then written as 
\begin{equation}\label{eq:6d-top-action}
    \begin{split}
        S_{\text{6d}}^{\scriptscriptstyle\text{Top}}\,\supset\, \int_{\mathcal{M}_{6}}\,&\,f_{1}^{(1)}\,\wedge\, \left(d\widetilde{f}_{4}^{\,(9)} - f_{2}^{(7)}\wedge h_{3}^{(3)} + f_{5}^{(7)}\,\wedge\,h_{0}^{(3)}\right)
        \\
        \, -\int_{\mathcal{M}_{6}}\,&\,\widetilde{f}_{1}^{(3)}\,\wedge\,\left( \, d\widetilde{f}_{4}^{\,(7)} + f_{5}^{(5)}\wedge h_{0}^{(3)} - f_{2}^{(5)} \wedge h_{3}^{(3)} \, \right)
        \\
    +\int_{\mathcal{M}_{6}}\,&\, h_{1}^{(3)}\,\wedge\,\left( \, d\widetilde{h}_{4}^{(7)} +  f_{5}^{(5)}\wedge \widetilde{f}_{0}^{\,(3)} - f_{2}^{(5)} \wedge \widetilde{f}_{3}^{\,(3)} \, \right)\,
    \\
    +\int_{\mathcal{M}_{6}}\,&\,  ( \,  h_{1}^{(3)}\wedge f_{1}^{(1)}\wedge \widetilde{f}_{4}^{\,(7)} \, ) \,.
    \end{split}
\end{equation}
We claim that the above action is capable of defining the symmetry topological operators whose lower-dimensional Page charges are given in (\ref{eq:def-F49}--\ref{eq:def-F11}). To demonstrate this, we first consider the equations of motion for $f_{1}^{(1)}$, $\widetilde{f}_{1}^{\,(3)}$, and $h_{1}^{(3)}$, which are respectively given by
\begin{equation}
    \begin{split}
        d\widetilde{f}_{4}^{\,(9)} - f_{2}^{(7)}\wedge h_{3}^{(3)} + f_{5}^{(7)}\,\wedge\,h_{0}^{(3)} - h_{1}^{(3)}\wedge \widetilde{f}_{4}^{\,(7)}\,&=\,0\,,
        \\
        d\widetilde{f}_{4}^{\,(7)} + f_{5}^{(5)}\wedge h_{0}^{(3)} - f_{2}^{(5)} \wedge h_{3}^{(3)} \,&=\,0\,,
        \\
        d\widetilde{h}_{4}^{(7)} +  f_{5}^{(5)}\wedge \widetilde{f}_{0}^{\,(3)} - f_{2}^{(5)} \wedge \widetilde{f}_{3}^{\,(3)} + f_{1}^{(1)}\wedge\widetilde{f}_{4}^{\,(7)}\,&=\,0\,.
    \end{split}
\end{equation}
Now using the definitions in (\ref{eq:def-F49}--\ref{eq:def-H47}), we can rewrite the above equations as
\begin{equation}
    \begin{split}
        dF_{4}^{(9)}\,&=\, d\widetilde{F}_{4}^{(9)}\,- \, h_{1}^{(3)}\wedge \widetilde{f}_{4}^{\,(7)}\,=\,0\,,
        \\
        dF_{4}^{(7)}\,&=\, 0\,,
        \\
        dH_{4}^{(7)}\,&=\, d\widetilde{H}_{4}^{(7)}\,  + \, f_{1}^{(1)}\wedge\widetilde{f}_{4}^{\,(7)}\,=\,0\,.
    \end{split}
\end{equation}
These equations also serve as definitions for the fields $\widetilde{F}_{4}^{(9)}$ and $\widetilde{H}_{4}^{(7)}$. Furthermore, the equations of motion of $\widetilde{f}_{4}^{\,(9)}$, $\widetilde{f}_{4}^{\,(7)}$, and $\widetilde{h}_{4}^{\,(7)}$ are given as 
\begin{equation}
    \begin{split}
        df_{1}^{(1)}\,&=\, dF_{1}^{(1)}\,=\,0\,,
        \\
        d\widetilde{F}_{1}^{\,(3)} + h_{1}^{(3)}\wedge \widetilde{f}_{1}^{\,(1)}\,&=\, dF_{1}^{(3)}\,=\,0\,,
        \\
        dh_{1}^{(3)}\,&=\, dH_{1}^{(3)}\,=\,0\,.
    \end{split}
\end{equation}
Here, we defined $\widetilde{F}_{1}^{(3)}:= \widetilde{f}_{1}^{\,(3)}$. With the above considerations in mind, we can rewrite the 6d auxiliary action as
\begin{equation}\label{eq:6d-top-KE-terms(2)}
 \begin{split}
        S_{\text{6d}}^{\scriptscriptstyle\text{Top}} \, \supset\,&\int_{\mathcal{M}_{6}}\, \left(\,F_{1}^{(1)}\,\wedge\,d\widetilde{F}_{4}^{(9)}\,-\,\widetilde{F}_{1}^{(3)}\,\wedge\,dF_{4}^{(7)}\,+\, H_{1}^{(3)}\,\wedge\,d\widetilde{H}_{4}^{(7)} \,\right)\,
        \\
        \,&\,+\,\int_{\mathcal{M}_{6}}\, \left(\,  H_{1}^{(3)}\wedge F_{1}^{(1)}\wedge \widetilde{f}_{4}^{\,(7)}  \,\right)\,.
 \end{split}
\end{equation}
With this more compact form of the 6d auxiliary action, we are ready to extract the lower-dimensional Page charges and, consequently, the symmetry topological operators. The key idea lies in using the last term, $H_{1}^{(3)} \wedge F_{1}^{(1)} \wedge \widetilde{f}_{4}^{\,(7)}$, to refine the operators constructed from the BF terms to form the correct symmetry topological operators. This is motivated by a similar strategy used in our work in \cite{Najjar:2024vmm}.

From the BF terms one observes that there are three different operators need to be refined to get the right symmetry operators, namely
\begin{equation}
    \exp(i\frac{\alpha_{\scriptscriptstyle(1)}}{2\pi}\int_{\Sigma_{1}} \widetilde{F}_{1}^{(3)})\,, \quad \exp(i\frac{\alpha_{\scriptscriptstyle(2)}}{2\pi}\int_{\Sigma_{4}} \widetilde{F}_{4}^{(9)})\,,\quad \exp(i\frac{\alpha_{\scriptscriptstyle(3)}}{2\pi}\int_{\Sigma_{4}} \widetilde{H}_{4}^{(7)})\,.
\end{equation}
The aim is to refine the above integrands such that we are able to construct the currents defined in (\ref{eq:def-F49}), (\ref{eq:def-H47}), and (\ref{eq:def-H13}). Let us examine one example in detail; the others follow from a similar treatment. In particular, we would like to refine the operators associated with $\widetilde{F}_{4}^{(9)}$. First, let us define the map
\begin{equation}
    \phi_{(1)}^{\ast} \ : \ \Omega^{\bullet} (\mathcal{M}_{6}) \longrightarrow \Omega^{\bullet} (\Sigma_{5}) \,.
\end{equation}
The lower-dimensional Page charge can then be written as
\begin{equation}
   d Q_{4}^{\scriptscriptstyle\text{D(-1)}}\,=\,\int_{\Sigma_{5}} \phi_{(1)}^{\ast}(d\widetilde{F}_{4}^{(9)}) - \phi_{(1)}^{\ast}\,( h_{1}^{(3)}\wedge \widetilde{f}_{4}^{\,(7)} \,)\, =\,\int_{\Sigma_{5}} \phi_{(1)}^{\ast}(d{F}_{4}^{(9)}) =\,0. 
\end{equation}
Here, the last term of (\ref{eq:6d-top-KE-terms(2)}) gives $h_{1}^{(3)}\wedge\widetilde{f}_{4}^{\,(7)}$ on $\Sigma_{5}$ such that the combination agrees with the $F_{4}^{(9)}$ as given in (\ref{eq:def-F49}). Now, assuming $\partial\Sigma_{5}=\Sigma_{4}$ and omitting the map $\phi_{(1)}^{\ast}$, the Page charge can be expressed as
\begin{equation}
     Q_{4}^{\scriptscriptstyle\text{D(-1)}}\,=\, \int_{\Sigma_{4}} {F}_{4}^{(9)}\,, 
\end{equation}
which gives (\ref{eq:sym-top-op-P9}) upon applying the exponential map. 

Having we demonstrated that the above 6d auxiliary action gives the symmetry topological operators of interests, we can reexpress it in a more compact way as
\begin{equation}\label{eq:6d-top-KE-terms(3)}
 \begin{split}
        S_{\text{6d}}^{\scriptscriptstyle\text{Top}} \, \supset\,&\int_{\mathcal{M}_{6}}\, \left(\,F_{1}^{(1)}\,\wedge\,d\widetilde{F}_{4}^{(9)}\,-\,\widetilde{F}_{1}^{\,(3)}\,\wedge\,dF_{4}^{(7)}\,+\, H_{1}^{(3)}\,\wedge\,d{H}_{4}^{(7)} \,\right)\,.
 \end{split}
\end{equation}
Here, the term $H_{1}^{(3)}\wedge F_{1}^{(1)}\wedge \widetilde{f}_{4}^{\,(7)}$  is absorbed into the last term $H_{1}^{(3)}\,\wedge\,d{H}_{4}^{(7)}$. Alternatively, it could have been absorbed into $F_{1}^{(1)}\wedge d\widetilde{F}_{4}^{(9)}$. Both approaches yield equivalent results, and the choice is motivated by expressing the action in a compact form while keeping in mind the refinement discussed above.

From the above 6d topological action, the following 5d bulk SymTFT can be obtained, 
\begin{equation}\label{eq:5d-SymTFT-action}
    S_{\scriptscriptstyle\text{5d}}^{\scriptscriptstyle\text{SymTFT}} \, \supset\,\frac{1}{(2\pi)^{2}}\,\int_{\mathcal{M}_{5}}\,\left( F_{1}^{(1)}\,\wedge\,\widetilde{F}_{4}^{(9)}\,-\,\widetilde{F}_{1}^{\,(3)}\,\wedge\,F_{4}^{(7)}\,+\, H_{1}^{(3)}\,\wedge\,H_{4}^{(7)}\right)\,.
\end{equation}
Here, we have introduced the prefactor $\frac{1}{(2\pi)^{2}}$ to align with the conventions used in \cite{Najjar:2024vmm}.

Before concluding this section, let us comment on the structure of the BF action above, as discussed in \cite[App.B]{Najjar:2024vmm}. In the literature, the BF terms for continuous symmetries, schematically written as
\begin{equation}
    \frac{1}{(2\pi)^{2}}\,\int_{\mathcal{M}_{d+1}}\,F_{p}\,\wedge\,\widetilde{F}_{d-p+1}\,,
\end{equation}
have two distinct interpretations:
\begin{itemize}
\item In \cite{Brennan:2024fgj,Antinucci:2024zjp}, the field $\widetilde{F}_{d-p+1}$ is interpreted as an $\R$-gauge field, with an additional condition imposed on the quantization of periods. Note that \cite{Gagliano:2024off} provides a (type IIA/B) top-down approach that aligns with \cite{Brennan:2024fgj,Antinucci:2024zjp}. Such an approach and its application to the current work deserve further investigation, which is beyond the scope of this paper.
\item In 
\cite{Apruzzi:2024htg}, the field $\widetilde{F}_{d-p+1}$ is interpreted as another flux field, which is automatically closed and quantized.
\end{itemize}
However, in \cite{Najjar:2024vmm} as well as in this work, the fields $\widetilde{F}_{d-p+1}$ are quantized but not closed. Due to the refinement arguments presented above, the SymTFT action includes an additional term that updates $\widetilde{F}_{d-p+1}$ to a closed and quantized flux. As a result, our work naturally lines up with the proposal in \cite{Apruzzi:2024htg}.

\paragraph{The origin of $\theta_{\text{YM}}$ and instanton number.}

The existence of a continuous $(-1)$-form symmetry implies the presence of a continuous and compact parameter in the effective field theory. In the case at hand, we have three distinct compact parameters in the holographic 4d $\mathcal{N}=1$ KS set-up. Here, we aim to determine the $\theta_{\text{YM}}$ parameter of the 4d $\mathcal{N}=1$ $SU(M)$ theory. To achieve this, we follow the strategy outlined in \cite{Najjar:2024vmm}, which involves identifying the bulk inflow term that corrects the anomaly between shifting $\theta_{\text{YM}}$ by $2\pi$ and the background gauge field $B_{2}^{\mathrm{e}}$ of the electric $\mathbb{Z}_N^{\scriptscriptstyle [1,\mathrm{e}]}$ 1-form symmetry. Schematically, the partition functions of the theory at $\theta_{\text{YM}}+2\pi $ and $\theta_{\text{YM}}$ differs by \cite{Kapustin:2014gua, Gaiotto:2014kfa, Gaiotto:2017yup, Gaiotto:2017tne, Cordova:2019uob},
\begin{equation}\label{eq:Zdifferby2pi-Anomaly}
    \frac{Z[\theta_{\text{YM}}+2\pi,B_{2}^{\mathrm{e}}]}{Z[\theta_{\text{YM}},B_{2}^{\mathrm{e}}]}    = \exp{2\pi i \,\frac{M-1}{M}\int_{M_4}\mathcal{P}(B_{2}^{\mathrm{e}})}\,.
\end{equation}
Here, 
\begin{equation}
    \mathcal{P} \ : \ H^2 (M_4, \Z/ M\Z) \longrightarrow H^4 (M_4, \Z/ 2M\Z)    
\end{equation}
is the Pontryagin square operation. In the present case, $\mathcal{P}(B_{2}^{\mathrm{e}}) \mod M$ reduces to $ B_{2}^{\mathrm{e}} \smile B_{2}^{\mathrm{e}} $. More precisely, it holds that $\frac{1}{2} B_{2}^{\mathrm{e}} \smile B_{2}^{\mathrm{e}} \in H^4 (M_4, \mathbb{Z}/ 2M\mathbb{Z})$. If $M$ is odd, $\mathcal{P}(B_{2}^{\mathrm{e}}) = B_{2}^{\mathrm{e}} \smile B_{2}^{\mathrm{e}} \in H^4 (M_4, \mathbb{Z}/ M\mathbb{Z})$. However, if $M$ is even, the explicit form of $\mathcal{P}(B_{2}^{\mathrm{e}})$ involves the cup and cup-1 products. For simplicity, we neglect these subtleties and use the fact that $\mathcal{P}(B_{2}^{\mathrm{e}}) \mod M$ equals $B_{2}^{\mathrm{e}} \smile B_{2}^{\mathrm{e}}$ for both odd and even $M$.

The above anomalous behaviour \eqref{eq:Zdifferby2pi-Anomaly} can be corrected by the bulk inflow via the term \cite[(2.5)]{Cordova:2019uob}
\begin{equation}\label{eq:F1B2B2-term}
    \frac{M-1}{2M}\,\, \frac{F_1^{\mathrm{e}}}{2\pi}\,\smile \mathcal{P}(B_{2}^{\mathrm{e}})\,.
\end{equation}
Here, $F_1^{\mathrm{e}}$ is the curvature of $\theta_{\text{YM}}$ as $\theta_{\text{YM}}$ is promoted to a periodic scalar.

Let us now turn to apply the above algorithm to the $SU(M)$ SYM theory. First, we note that the electric $B_{2}^{\mathrm{e}}$ is originated from the type IIB NS-NS 3-form $H_{3}=dB_{2}$, see, e.g., \cite{Apruzzi:2021phx}. This observation and the structure of the inflow indicates that we should consider the topological type IIB CS-terms. Second, we should consider the consistent truncation of the type IIB supergravity on the $T^{1,1}$ conifold to 5d theory, as presented in \cite{Cassani:2010na,Bena:2010pr,Schon:2006kz}. In particular, consider the truncation of the $H_{3}\wedge F_{3}\wedge F_{5}$ term as in \cite[(3.24)]{Cassani:2010na}, yielding
\begin{equation}\label{eq:origin-of-dcb2b2}
\begin{split}
      -\, \int_{T^{1,1}} \, \widetilde{F}_{5}\,\wedge\, H_{3}\,\wedge\, \widetilde{F}_{3}\, &\,\supset\,-\frac{1}{2}\,\int_{\mathcal{M}_{6}} \, f_{2}^{\Phi}\,\wedge\, db_{2}^{(2)}\,\wedge (\,Dc^{\Phi} - c_{0}Db^{\Phi})  \, 
       \\
       \,&\,\supset\, -\frac{1}{2}\,\int_{\mathcal{M}_{6}}\,\, M\,b_{2}^{(2)}\,\wedge\, db_{2}^{(2)}\,\wedge\,dc^{\Phi}\,,
\end{split}
\end{equation}
with, $h_{3}^{(3)}=db_{2}^{(2)}$, $h_{1}^{(3)} = db^{\Phi}=db_{0}^{(2)}$, $Dc^{\Phi}=dc^{\Phi}- q A$ (with $A$ being the $U(1)$ R-symmetry gauge field), $Db^{\Phi} = db^{\Phi}-pA$, and $f_{2}^{\Phi}\supset \frac{M}{2}b_{2}$. For further details, the reader is invited to consider the original reference \cite{Cassani:2010na}.

In fact, the above term coincide with the $-h_{3}^{(3)}\wedge \widetilde{f}_{1}^{\,(3)}\wedge f_{2}^{(5)}$ term in (\ref{eq:6d-top-action}) when employ the following identifications (up to numerical factors),
\begin{equation}
   h_{3}^{(3)} = db_{2}^{(2)}=dB_{2}^{\mathrm{e}}\,,\qquad f_{2}^{(5)}\,\sim\, f_{2}^{\Phi}\,.
\end{equation}
The aforementioned term is already included in the second line of (\ref{eq:6d-top-action}). In particular, we have
\begin{equation}
   \begin{split}
      {f}_{1}^{(3)}\,\wedge\, f_{2}^{(5)} \wedge h_{3}^{(3)}  \,&\, \, \sim \,\, -dc^{\Phi}\,\wedge \,  \frac{M}{2}B_{2}^{\mathrm{e}} \wedge dB_{2}^{\mathrm{e}} \, \,.
   \end{split}
\end{equation}
Thus, the 5d SymTFT action in (\ref{eq:5d-SymTFT-action}) already contains the inflow (\ref{eq:F1B2B2-term}). Therefore, the above term does not appear separately in our 5d SymTFT. Now by matching this expression with (\ref{eq:F1B2B2-term}) and refining $\widetilde{F}_{1}^{(3)}$ to $F_{1}^{(3)}$, we identify
\begin{equation}\label{eq:F1(3)-dthetaYM}
    {F}_{1}^{(3)}\,\sim\,d\theta_{\text{YM}}\,.
\end{equation}
Hence, $\widetilde{F}_{1}^{(3)}$ is related to the curvature of $\theta_{\text{YM}}$ up to a shift as can be seen from (\ref{eq:origin-of-dcb2b2}). Note that, a similar identification appears in \cite[App.C]{Apruzzi:2021phx}.

From this analysis, we conclude that the $\theta_{\text{YM}}$ parameter originates from the type IIB RR 3-form $\widetilde{F}_{3}$. In the gravitational bulk ${M}_{5}$, the corresponding compact scalar operator arises from D1-branes wrapping the $S^{2}\subset T^{1,1}$. Moreover, $F_{4}^{(7)}$ is related to the Chern-Weil current, corresponding to the instanton density of the $SU(M)$ gauge theory. In particular, we have the following identifications:

\begin{equation}\label{eq:F4(7)-trFF}
  \frac{1}{2\pi}\,  F_{4}^{(7)}\quad \xleftrightarrow[]{\,\,\text{identify on $\mathcal{B}^{\scriptscriptstyle\text{phys}}$}\,\,}\quad \begin{cases}
       \frac{1}{8\pi^{2}}\,\tr(F\wedge F)\,,\qquad &\text{trivial $B_{2}^{\mathrm{e}}$}\,, \\
       \frac{1}{8\pi^{2}}\,\tr{(\widetilde{F}_{2}^{\mk{u}(M)}-B_{2}^{\mathrm{e}})^{2}}\,,\qquad &\text{non-trivial $B_{2}^{\mathrm{e}}$}\,.
   \end{cases}
\end{equation}
Here,  we adopt standard conventions from the literature, e.g., \cite{Gaiotto:2017yup,Cordova:2019uob}, where the $\mk{su}(M)$-gauge curvature $F_{2}^{\mk{su}(M)}$ is lifted to a $\mk{u}(M)$-gauge curvature $\widetilde{F}_{2}^{\mk{u}(M)}$ to introduce the electric 1-form gauge field $B_{2}^{\mathrm{e}}$, i.e.,
\begin{equation}
     F_{2}^{\mk{su}(M)} \ \mapsto \ \widetilde{F}_{2}^{\mk{u}(M)} - B_{2}^{\mathrm{e}}\,,
\end{equation}
subject to the condition
\begin{equation}
    \tr(\widetilde{F}_{2}^{\mk{u}(M)}) \,=\, M\,B_{2}^{\mathrm{e}}\,.
\end{equation}
For brevity, we omit the superscript over the gauge curvature in the subsequent discussion.

\subsection{Modified instanton sum and 4-group structure}

In this section, we demonstrate that a consistent decomposition can be defined for the holographic dual gauge theory of the Klebanov-Strassler solution. Specifically, we show that the pure 4d $\N=1$ $SU(M)$ gauge theory can be consistently coupled to a (hidden) TQFT sector, which includes discrete 2-form and 3-form symmetry. This coupling allows for a modified instanton sum and the emergence of a 4-group structure, in the sense discussed in \cite{Tanizaki:2019rbk}.

This construction parallels the approach taken in the context of geometric engineering, as detailed in \cite[Sec 4.4]{Najjar:2024vmm}. Consequently, the discussion in this subsection builds on those ideas, extending them to the holographic dual framework.

The TQFT sector that couple to the gauge theory can be revealed by considering the 5d bulk SymTFT theory given in (\ref{eq:5d-SymTFT-action}),
\begin{equation}\label{eq:the-5d-SymTFT}
\begin{split}
    S_{\scriptscriptstyle\text{5d}}^{\scriptscriptstyle\text{SymTFT}} \, &= \, \frac{1}{(2\pi)^{2}}  \int_{\mathcal{M}_{5}} \left( F_{1}^{(1)} \wedge \widetilde{F}_{4}^{(9)} - \widetilde{F}_{1}^{(3)} \wedge F_{4}^{(7)} + H_{1}^{(3)} \wedge H_{4}^{(7)} \right)\,.
\end{split}
\end{equation}

As demonstrated around (\ref{eq:F1(3)-dthetaYM}) and (\ref{eq:F4(7)-trFF}), the second term of the first line above gives the known gauge theory $\theta_{\text{YM}}$-term upon projecting the theory onto the physical boundary $\mathcal{B}^{\scriptscriptstyle\text{phys}}$. The projection can be achieved by defining a projection operator as done in \cite[App. B]{Najjar:2024vmm}. Without going into many details, we define a projection operator $\widetilde{\delta}$ as a restriction of the SymTFT field strengths and background gauge fields to the physical boundary $\mathcal{B}^{\scriptscriptstyle\text{phys}}$, 
\begin{equation}\label{eq:def-delta-operator}
    \widetilde{\delta}(\mathscr{F})\,\equiv \,\restr{\mathscr{F}}{\mathcal{B}^\text{phys}}\,.
\end{equation}
Here, $\mathscr{F}$ refer to any field appears in the bulk SymTFT. The operator $\widetilde{\delta}$ plays the role of mapping the SymTFT fields $\mathscr{F}$ to topological terms within the physical QFT, subject to the prescribed boundary conditions at the symmetry boundary $\mathcal{B}^{\scriptscriptstyle\text{sym}}$. Applying the projection operator to $\widetilde{F}_{1}^{(3)}\wedge F_{4}^{(7)}$, then via (\ref{eq:F4(7)-trFF}) we have the following two cases:
\begin{itemize}
    \item Trivial background field of the $\Z_{N}^{\scriptscriptstyle[1,\mathrm{e}]}$, i.e., $B_{2}^{\mathrm{e}}=0$,
\begin{equation}\label{eq:rep-F47-case-1}
  -\,\frac{1}{(2\pi)^{2}}  \widetilde{\delta}\left(\widetilde{F}_{1}^{(3)}\wedge F_{4}^{(7)} \right)\,=\, \frac{\theta_{\text{YM}}}{8\pi^{2}}\,\tr(F\,\wedge\,F)\,.
\end{equation}
\item Non-trivial background field of the $\Z_{N}^{\scriptscriptstyle[1,\mathrm{e}]}$, i.e., $B_{2}^{\mathrm{e}}\neq0$,
\begin{equation}\label{eq:rep-F47-case-2}
       -\,\frac{1}{(2\pi)^{2}}  \widetilde{\delta}\left(\widetilde{F}_{1}^{(3)}\wedge F_{4}^{(7)} \right)\,=\, \frac{\theta_{\text{YM}}}{8\pi^{2}}\,\tr{(\widetilde{F}_{2} - B_{2}^{\mathrm{e}})^{2}}\,.
\end{equation}
\end{itemize}

To consistently define the modified instanton sum and the 4-group structure, we must gauge specific finite subgroups of the two remaining continuous symmetries in the KS set-up, which we discuss next.

\paragraph{From continuous to discrete symmetries.} The BF terms associated with continuous symmetries in this work take the flux-flux form, i.e.,
\begin{equation}
    \frac{1}{(2\pi)^{2}}\,\int_{\mathcal{M}_{d+1}}\, F_{p+2}\wedge H_{d-p-1}\,,
\end{equation}
which suggests the presence of a continuous $p$-form symmetry or $(d-p-3)$-form symmetry. This structure aligns with results from a field-theoretic perspective \cite{Apruzzi:2024htg} and from an M-theory perspective \cite{Najjar:2024vmm}. 

Here, we would like to gauge particular finite subgroups of continuous symmetries via determining the realization of SymTFT field strengths on the physical boundary $\mathcal{B}^{\scriptscriptstyle\text{phys}}$. In particular, let us apply the operator $\widetilde{\delta}$ given in (\ref{eq:def-delta-operator}) on the above flux-flux term,
\begin{equation}\label{eq:F-H-flux-flux-SymTFT}
\frac{1}{(2\pi)^{2}}\,\widetilde{\delta}\left(\int_{\mathcal{M}_{d+1}}\, F_{p+2}\wedge H_{d-p-1}\right) \,=\,\frac{1}{(2\pi)^{2}}\, \int_{M_{d}}\, \widetilde{\delta}(F_{p+2})\,\wedge\,\widetilde{\delta}(H_{d-p-1})\,.
\end{equation}
In this context, there are two possible realizations of $\widetilde{\delta}(\text{$G$-flux})$ on $\mathcal{B}^{\scriptscriptstyle\text{phys}}$. Schematically, these are given by
\begin{equation}\label{eq:2-options-for-flux}
 \widetilde{\delta}(\text{$G_{p}$-flux}) \,:\,\,\,
\begin{cases}
\text{$G_{p}$-flux}: \qquad \  &\widetilde{\delta}(G_{p})=\widetilde{\delta}(dC_{p-1})\,,
\\
\text{$C_{p}$-gauge field}:\qquad\ & \widetilde{\delta}(G_{p})=C_{p} \,.
\end{cases}
\end{equation}
In our notation, we use lower-case Latin letters to denote the gauge fields of the gauged symmetry, while upper-case Latin letters are reserved for background fields.

When describing a discrete $\Z_{N}^{\scriptscriptstyle[p]}$ $p$-form symmetry, we use a pair of $U(1)$-valued gauge fields $(C_{p+1}, C_{p})$, subject to the constraint
\begin{equation}\label{eq:NC=dC}
    N C_{p+1}\,=\,dC_{p}\,.
\end{equation}
Using such conventions justify the second case of (\ref{eq:2-options-for-flux}).

Applying the above classification to the BF-term in (\ref{eq:F-H-flux-flux-SymTFT}), we arrive at various distinct possibilities. First, let us assume that $-1\leq p\leq d-2$, then two options follow from the first case of (\ref{eq:2-options-for-flux}):
\begin{itemize}
\item Gauging $\Z_{N}$ $(d-p-3)$-form symmetry,
\begin{equation}
    \begin{split}
\frac{1}{(2\pi)^{2}}\, \int_{M_{d}}\, \widetilde{\delta}(F_{p+2})\,\wedge\,\widetilde{\delta}(H_{d-p-1}) \,&=\, \frac{1}{(2\pi)^{2}}\, \int_{M_{d}}\, \widetilde{\delta}(2\pi N\,dA_{p+1})\,\wedge\,\widetilde{\delta}(dc_{d-p-2})
\\
\,&=\,\frac{N}{2\pi}\,\int_{M_{d}}\, A_{p+1}\wedge dc_{d-p-2}\,.
\end{split}
\end{equation}
Here, $c_{d-p-2}$ represents a spacetime gauge field associated with the discrete $\Z_{N}^{\scriptscriptstyle[d-p-3]}$ symmetry, while $A_{p+1}$ functions as a background gauge field used to construct the corresponding symmetry topological operator.

For example, the case of $d=4$ and $p=-1$ gives a gauged 2-form symmetry with the following 4d action,
\begin{equation}
    \frac{N}{2\pi}\,\int\,A_{0}\,\wedge\,dc_{3}\,.
\end{equation}
In this paper, we also refer to 0-form background fields via the Greek latter $\chi$. 

\item Gauging $\Z_{N}$ $p$-form symmetry,
\begin{equation}
\begin{split}
\frac{1}{(2\pi)^{2}}\, \int_{M_{d}}\, \widetilde{\delta}(F_{p+2})\,\wedge\,\widetilde{\delta}(H_{d-p-1}) \,&=\, \frac{1}{(2\pi)^{2}}\, \int_{M_{d}}\, \widetilde{\delta}(da_{p+1})\,\wedge\,\widetilde{\delta}(2\pi N\,dC_{d-p-2})
\\
\,&=\,\frac{N}{2\pi}\,\int_{M_{d}}\, a_{p+1}\wedge dC_{d-p-2}
\end{split}
\end{equation}
In this case, $a_{p+1}$ is the spacetime gauge field and $C_{d-p-2}$ is a background gauge field.

For the specific case of $d=4$ and $p=-1$, this corresponds to a gauged $(-1)$-form symmetry, described by the term 
\begin{equation}
    \frac{N}{2\pi}\,\int_{M_{4}}\,a_{0}\,\wedge\,dC_{3}\,.
\end{equation}
\end{itemize}
The above two scenarios provide a natural mechanism of gauging a $\Z_{N}$ subgroup of the continuous $U(1)^{[d-p-3]}$ and $U(1)^{[p]}$ symmetries, respectively.  However, this gauging cannot be performed simultaneously, and a choice must be made between the two.  Once a particular choice is adopted, and given that the discrete gauged symmetries are not dual to each other, a natural question arises: how can one change the polarization and gauge their dual discrete symmetries? This question motivates us to explore the second case of (\ref{eq:2-options-for-flux}), where the field strength is treated as a gauge field on the physical boundary $\mathcal{B}_{d}^{\scriptscriptstyle\text{phys}}$. Furthermore, adopting such a perspective enable us to extend the possible values of $p$ beyond our previous assumption, $-1\leq p\leq d-2$.  

Consequently, two additional possibilities emerge from considering this second case, as we now describe:
\begin{itemize}
\item Gauging $\Z_{N}$ $(d-p-2)$-form symmetry,
\begin{equation}
\begin{split}
    \frac{1}{(2\pi)^{2}} \int_{M_{d}}\widetilde{\delta}(F_{p+2}) \,\wedge\, \widetilde{\delta}(H_{d-p-1})\,&=\,\frac{N}{2\pi}\, \int_{M_{d}}\, (\widetilde{\delta}(F_{p+2}))_{p+1}\,\wedge\,  c_{d-p-1}\,.
\end{split}
\end{equation}
In this expression, we apply the second case of (\ref{eq:2-options-for-flux}) to $\widetilde{\delta}(H_{d-p-1})$, while the $(p+1)$-form representation of $F_{p+2}$, i.e., $(\widetilde{\delta}(F_{p+2}))_{p+1}$ depends on the value of $p$. Specifically, $c_{d-p-2}$  denotes a dynamical $U(1)$-valued gauge field according to (\ref{eq:NC=dC}). The background gauge field is provided by $(\widetilde{\delta}(F_{p+2}))_{p+1}$. 

For the case of $d=4$ and $p=-1$, then $(\widetilde{\delta}(F_{p+2}))_{p+1}$ gives a background 0-form field, which follows from the first case of (\ref{eq:2-options-for-flux}), i.e.,
\begin{equation}
  \frac{N}{2\pi}\, \int_{M_{d}}\, (\widetilde{\delta}(F_{1}=dA_{0}))_{0}\,\wedge \, c_{4}\,=\,  \frac{N}{2\pi}\,\int_{M_{4}}\, A_{0} \,\wedge\, c_{4}\,.
\end{equation}

\item Gauging $\Z_{N}$ $(p+1)$-form symmetry,
\begin{equation}
\begin{split}
        \frac{1}{(2\pi)^{2}}\, \int_{M_{d}}\, \widetilde{\delta}(F_{p+2})\,\wedge\,\widetilde{\delta}(H_{d-p-1})\,&=\, \frac{N}{2\pi}\, \int_{M_{d}}\, a_{p+2}\,\wedge\,(\widetilde{\delta}(H_{d-p-1}))_{d-p-2}\,.
\end{split}
\end{equation}
In this alternative scenario, we apply the second case of (\ref{eq:2-options-for-flux}) to $\widetilde{\delta}(F_{p+2})$ and again the representation of $(\widetilde{\delta}(H_{d-p-1}))_{d-p-2}$ depends on the value of $p$.

For our illustrative example with $d=4$ and $p=-1$, the component $(\widetilde{\delta}(H_{d-p-1}))_{d-p-2}$ corresponds to a 3-form field strength, which explicitly takes the form $dC_{2}$, i.e. we have
\begin{equation}
\frac{N}{2\pi}\, \int_{M_{4}}\, a_{1}\wedge dC_{2}\,.
\end{equation}
Notably, in this example, $(\widetilde{\delta}(H_{d-p-1}))_{d-p-2}$ follows from the first case of (\ref{eq:2-options-for-flux}), where the flux is obtained from a lower-degree form field.

\end{itemize}

\paragraph{Modified instanton sum.}

To get the modified instanton sum, we need to couple the gauge theory to a gauged discrete $\Z_{p}^{[2]}$ 2-form symmetry \cite{Seiberg:2010qd,Tanizaki:2019rbk,Najjar:2024vmm}. To achieve that, we proceed with the following prescription:   

\begin{itemize}
\item Introduce the $\theta_{\text{YM}}$ along with a background field $\chi^{(1)}$ by taking
\begin{equation}
    \widetilde{\delta}\left(\widetilde{F}_{1}^{(3)}\right)\,=\,\theta_{\text{YM}}\,+\,\chi^{(1)}\,.
\end{equation}
\item Gauge a $\Z_{p}^{[2]}$ subgroup of the continuous $U(1)^{[2]}$ 2-form symmetry. Since there are two continuous 2-form symmetries, then a choice need to be taken. Without loss of generality, we take
\begin{equation}
    \frac{1}{2\pi}\widetilde{\delta}\left(H_{4}^{(7)}\right)\,=\,dc_{3}\,,
\end{equation}
where $c_{3}$ being the gauge field for the continuous 2-form symmetry generated by $H_{1}^{(3)}$. The projection of $H_{1}^{(3)}$ is taken as,
\begin{equation}
    \widetilde{\delta}\left(H_{1}^{(3)}\right)\,=\,\theta^{(2)}\,+\, p\,\chi^{(2)}\,, \quad p\in \Z\,.
\end{equation}
Here, $\chi^{(2)}$ is another 0-form Lagrange multiplier. Its equation of motion implies that $p\,dc_{3}=0$, which reflects the fact that we gauged a $\Z_{p}$ subgroup.
\end{itemize}

Following the above prescription, the 4d TQFT theory is then given by
\begin{equation}\label{eq:Top-action-mod-(1)}
 S_{\text{TQFT}}^{SU(N)/\Z_{p}^{[2]}}\,=\, \int_{M_{4}}\,\,\left[   \frac{\theta_{\text{YM}} + \chi^{(1)}}{8\pi^{2}}\,\tr(F\,\wedge\,F)\,+\, \frac{\theta^{(2)}+p\,\chi^{(2)}}{2\pi}\,dc_{3}\,\right]\,.
\end{equation}
The equation of motion for $\chi^{(1)}$ implies
\begin{equation}
    \frac{1}{8\pi^{2}}\int_{M_{4}}\, \tr(F\,\wedge\,F)\, =\, 0\,,
\end{equation}
setting the instanton number to zero. To have non-trivial result, we either set $\chi^{(1)}$ to zero from the beginning, or we demand
\begin{equation}
    \chi^{(1)} \,=\, - \, \chi^{(2)} \,:=\, \chi \,.
\end{equation}
For the latter case, we arrive at the modified instanton sum \cite{Seiberg:2010qd,Tanizaki:2019rbk},
\begin{equation}
      \frac{1}{8\pi^{2}}\int_{M_{4}}\, \tr(F\,\wedge\,F) \,=\, p\,\int_{M_{4}}\,dc_{3}\,.
\end{equation}
Since $\int_{M_{4}}\,dc_{3}\in \Z$, then the instanton number is a multiple of $p$. 

The topological action (\ref{eq:Top-action-mod-(1)}) is now written as
\begin{equation}\label{eq:Top-action-mod-(2)}
\begin{split}
   S_{\text{TQFT}}^{SU(N)/\Z_{p}^{[2]}}\,=\,  &\int_{M_{4}}\,\,\left[   \frac{\theta_{\text{YM}}}{8\pi^{2}}\,\tr(F\,\wedge\,F)\,+\, \frac{\theta^{(2)}}{2\pi}\,dc_{3}\,\right]
     \\
 &+\int_{M_{4}}\,\,\chi\,\wedge\,\left[\,\frac{1}{8\pi^{2}}\tr(F\wedge\ F)- \frac{p}{2\pi}\,dc_{3}\,\right]\,.
\end{split}
\end{equation}

\paragraph{The 4-group structure.}When a non-trivial background field for the electric 1-form symmetry is turned on, the projection of the $\widetilde{F}_{1}^{(3)}\wedge F_{4}^{(7)}$ term is determined by (\ref{eq:rep-F47-case-2})—equivalently, it follows from the second case of (\ref{eq:F4(7)-trFF}). Consequently, one expects the 4d TQFT action to take the form
\begin{equation}
\begin{split}
   S_{\text{TQFT}}^{SU(N)/\Z_{N}^{[1]}\times\Z_{p}^{[2]}}\,=\,  &\int_{M_{4}}\,\,\left[   \frac{\theta_{\text{YM}}}{8\pi^{2}}\,\tr(\widetilde{F}-B_{2}^{\mathrm{e}})^{2}\,+\, \frac{\theta^{(2)}}{2\pi}\,dc_{3}\,\right]
     \\
 &+\int_{M_{4}}\,\,\chi\,\wedge\,\left[\,\frac{1}{8\pi^{2}}\tr(\widetilde{F}-B_{2}^{\mathrm{e}})^{2}- \frac{p}{2\pi}\,dc_{3}\,\right]\,.
\end{split}
\end{equation}
However, as already pointed out in \cite{Tanizaki:2019rbk}, the above action is not invariant under the gauge transformation
\begin{equation}\label{eq:b2-to-b2-l1}
    B_{2}^{\mathrm{e}}\to B_{2}^{\mathrm{e}}+d\Lambda_{1}\,,
\end{equation}
where $\Lambda_{1}$ is a 1-form gauge parameter. This lack of gauge invariance suggests the need for a more refined structure.

To restore gauge invariance under this transformation, one must gauge a discrete 3-form symmetry and enforce a 4-group structure. The 4-group is characterized by the following twisted product
\begin{equation}\label{4-group-structure}
   \left( \Z_{N}^{\scriptscriptstyle [1,\mathrm{e}]} \times \Z_{p}^{\scriptscriptstyle [2]} \right)\  \widetilde{\times} \ \Z_{p}^{\scriptscriptstyle [3]} \,.
\end{equation}
Specifically, the gauge transformation of the $\Z_{p}^{\scriptscriptstyle[3]}$ gauge field is influenced by the transformation properties of the electric 1-form symmetry.

To derive this 4-group structure, we follow a similar prescription to that outlined in \cite[Sec 4.4]{Najjar:2024vmm}:
\begin{itemize}
    \item We gauge the electric 1-form symmetry. In particular, we impose the Neumann boundary condition on the SymTFT $B_{2}^{\mathrm{e}}$ gauge fields as 
\begin{equation}
    \widetilde{\delta}\left(B_{2}^{\mathrm{e}}\,\wedge\,B_{2}^{\mathrm{e}}\right)\,=\, b_{2}\,\wedge\,b_{2}\,.
\end{equation}
Consequently, the projection of the SymTFT $F_{4}^{(7)}$ field strength is now given as
\begin{equation}
    \frac{1}{2\pi}\,\widetilde{\delta}\left(F_{4}^{(7)}\right)\,=\, \tr{(\widetilde{F}_{2}-b_{2})^{2}}\,.
\end{equation}

\item We introduce and gauge a discrete 3-form symmetry, where the charged objects are non-dynamical domain walls. In the context of the AdS/CFT correspondence, these domain walls correspond to BPS branes. However, if we aim to preserve the KS solution, there are no apparent 3-dimensional defects charged under a genuine discrete 3-form symmetry that can be consistently incorporated into the KS background.

Given this restriction, the only viable alternative is to gauge a continuous symmetry, following the framework outlined in the discussion after (\ref{eq:2-options-for-flux}). 

In particular, we project the $F_{1}^{(1)}\wedge \widetilde{F}_{4}^{(9)}$ term of (\ref{eq:the-5d-SymTFT}) in the following way:
\begin{itemize}
    \item We introduce a third $\theta$-parameter and a third background 0-form via
\begin{equation}
\begin{split}
        \widetilde{\delta}\left(F_{1}^{(1)}\right)\,&=\,\widetilde{\delta}\left(d(\theta^{(3)}\,+ \, q\,\chi^{(3)})\right)\,
        \\
        &=\,\theta^{(3)}\,+ \, q\,\chi^{(3)}\,.
\end{split}
\end{equation} 

\item We project $\widetilde{F}_{4}^{(9)}$ according to the second case of (\ref{eq:2-options-for-flux}). Specifically, on the physical boundary $\widetilde{F}_{4}^{(9)}$ is considered as a 4-form gauge field, 
\begin{equation}
    \frac{1}{2\pi}\widetilde{\delta}\left(\widetilde{F}_{4}^{(9)}\right)\,=\, a_{4}\,.
\end{equation}
Here, the $U(1)$-valued gauge field $a_{4}$ is interpreted as the gauge field of the gauged $\Z_{q}^{[3]}$ 3-form symmetry. From (\ref{eq:NC=dC}), such a gauge field should locally satisfy (e.g. \cite{Tanizaki:2019rbk,Najjar:2024vmm})
\begin{equation}\label{eq:locally-qa4=da3}
   a_{4}\,=\,\frac{1}{q}\,da_{3}\quad \Rightarrow\quad q\,a_{4}\,=\,da_{3} \quad \text{(locally)}\,. 
\end{equation}
\end{itemize}
\end{itemize}
Applying the above prescription extends the TQFT action given in (\ref{eq:Top-action-mod-(2)}) to the following
\begin{equation}
\begin{split}
  &\int_{M_{4}}\,\,\left[   \frac{\theta_{\text{YM}}}{8\pi^{2}}\,\tr(\widetilde{F}-b_{2})^{2}\,+\, \frac{\theta^{(2)}}{2\pi}\,dc_{3}\, + \frac{\theta^{(3)}}{2\pi}\,a_{4} \right]
     \\
 &+\int_{M_{4}}\,\,\chi\,\wedge\,\left[\,\frac{1}{8\pi^{2}}\tr(\widetilde{F}-b_{2})^{2}- \frac{p}{2\pi}\,dc_{3}\,\right]\,+\int_{M_{4}}\,  \frac{q}{2\pi}\chi^{(3)}\wedge a_{4}\,.
\end{split}
\end{equation}
Invariance under the large gauge transformation associated with the gauge field of the gauged 2-form symmetry,  
\begin{equation}
    c_{3} \, \to \, c_{3}\,+\, \Lambda_{3}\,,
\end{equation}
implies additional constraints on the above topological action. In particular, having that we demand $a_{4}$ to transform as
\begin{equation}\label{eq:a4L3gauge}
    a_4 \ \mapsto \ a_4 + \dd \Lambda_{3} \,,
\end{equation}
then we shall take the following identification 
\begin{equation}
    \theta^{(2)} \, = \, - \,\theta^{(3)} \,:=\, \theta\,, \qquad 
     \chi^{(2)} \, =\, -\chi^{(3)}\,,\qquad p=q\,.
\end{equation}
As a result, the 4d TQFT action is given as
\begin{equation}
\begin{split}
  &\int_{M_{4}}\,\,\left[   \frac{\theta_{\text{YM}}}{8\pi^{2}}\,\tr(\widetilde{F}-b_{2})^{2}\,+\, \frac{\theta}{2\pi}\,(dc_{3}\, - \,a_{4}) \right]
     \\
 &+\int_{M_{4}}\,\,\chi\,\wedge\,\left[\,\frac{1}{8\pi^{2}}\tr(\widetilde{F}-b_{2})^{2}- \frac{p}{2\pi}\,(dc_{3} -  a_{4})\,\right]\,\,.
\end{split}
\end{equation}

To achieve invariance of the above action under the gauge transformation (\ref{eq:b2-to-b2-l1}), we demand two thing \cite{Tanizaki:2019rbk}:
\begin{itemize}
    \item We relax the local behaviour of $a_{4}$ given in (\ref{eq:locally-qa4=da3}) and instead demand that $a_{4}$ satisfies 
\begin{equation}
    \frac{p}{2\pi}\,a_{4}\,=\, \frac{1}{2\pi}\,da_{3} \,+\, \frac{N}{8\pi^{2}}\,b_{2}\wedge b_{2}\,.
\end{equation}

\item We take $a_{4}$ as given above to be invariant under the 1-form gauge transformation, therefore $a_{3}$ should transform as
\begin{equation}
    a_{3} \ \mapsto \ a_{3} - \frac{N}{2\pi}b_{2}\wedge \Lambda_{1} - \frac{N}{4\pi}\Lambda_{1}\wedge \dd \Lambda_{1}\,.
\end{equation}
This particular transformation of $a_{3}$ reflects the non-trivial 4-group structure given in (\ref{4-group-structure}).
\end{itemize}
As a result, we can rewrite the 4d TQFT action as the following
\begin{equation}
\begin{split}
 S_{\text{TQFT}}^{SU(N)/  \left( (\Z_{N}^{\scriptscriptstyle [1,\mathrm{e}]} \times \Z_{p}^{\scriptscriptstyle [2]})  \widetilde{\times}  \Z_{p}^{\scriptscriptstyle [3]}\right)} \,=\, &\int_{M_{4}}\,\,\left[   \frac{\theta_{\text{YM}}}{8\pi^{2}}\,\tr(\widetilde{F}-b_{2})^{2}\,+\, \frac{\theta}{2\pi}\,(dc_{3}\, - \,a_{4}) \right]
     \\
 &+\int_{M_{4}}\,\,\chi\,\wedge\,\left[\,\frac{1}{8\pi^{2}}\tr(\widetilde{F})^{2}- \frac{p}{2\pi}\,dc_{3} + \frac{1}{2\pi}\,da_{3}\,\right]\,\,.
\end{split}
\end{equation}

For further discussion and implications of the 4d action regarding the periodicity of $\theta$ parameters, the reader may consult \cite{Tanizaki:2019rbk}.

\paragraph{Decomposition.} Before concluding this section, let us lightly review the idea of decomposition and its relation to the 4-group structure. As mentioned earlier, the notion of decomposition was first discovered and revealed in \cite{Pantev:2005rh,Pantev:2005wj,Pantev:2005zs,Hellerman:2006zs}. Here, we present a light review of the concept of decomposition for invertible higher-form symmetries. 

For a $d$-dimensional QFT with a discrete $(d-1)$-form symmetry, decomposes into a sum of disjoint QFTs (often called "universes"), each labelled by an irreducible representation of the symmetry. For a finite $\Z_{K}^{[d-1]}$ $(d-1)$-form symmetry, the decomposition implies
\begin{equation}
  \text{QFT}\,=\,  \bigoplus_{p=0}^{K-1}\, \text{QFT}_{p}\,.
\end{equation}
It follows that the partition function and the Hilbert space decompose as
\begin{equation}
     Z \,=\, \sum_{p=0}^{K-1}\,Z_{p}\,,\qquad\,  \mathcal{H}\,=\,\bigoplus_{p=0}^{K-1}\, \mathcal{H}_{p}\,.
\end{equation}
Correlation functions between the different universes are zero \cite{Tanizaki:2019rbk,Sharpe:2022ene}. 

The discrete $\Z_{K}^{[d-1]}$ $(d-1)$-form symmetry implies the existence of heavy and non-dynamical domain-walls, i.e. $(d-1)$-dimensional defects, denoted by $\mathcal{D}_{q}(\Sigma_{d-1})$ with $q=0,1,\cdots, K-1$. The symmetry topological operators are given by 0-dimensional operators $\mathcal{O}_{p}(\wp)$, with $p=0,1,\cdots,K-1$. The action of the symmetry operator on the DWs defects is given by \cite{Gaiotto:2014kfa,Cherman:2021nox,Tanizaki:2019rbk}
\begin{equation}\label{eq:OD-action}
    \expval{\mathcal{O}_{p}(\wp)\,\mathcal{D}_{q}(\Sigma_{d-1})}\,=\,\exp{\frac{2\pi i}{K}\,pq\,\,\text{Link}(\Sigma_{d-1},\wp)}\,\expval{\mathcal{D}_{q}(\Sigma_{d-1})\,\mathcal{O}_{p}(\wp)}\,.
\end{equation}
Here, $\text{Link}(\Sigma_{d-1},\wp)$ is the linking number between $\Sigma_{d-1}$ and the point $\wp$.

Following \cite{Cherman:2021nox} and \cite[Sec.4]{Sharpe:2022ene}, one can define projectors $\Pi_{q}$ out of the local operators $\mathcal{O}_{p}(\wp)$ as
\begin{equation}
    \Pi_{q} = \frac{1}{K}\,\sum_{0}^{K-1}\,\exp(\frac{2\pi i }{K}\,pq)\, \mathcal{O}_{p}(\wp)\,
\end{equation}
Then it follows from (\ref{eq:OD-action}), that on each sides of a UV domain wall the projector operator differ according to
\begin{equation}
   \Pi_{q}\,\, \mathcal{D}_{\widetilde{q}}(\Sigma_{d-1})\,=\, \mathcal{D}_{\widetilde{q}} \,\, \Pi_{q+\widetilde{q}}\,,\qquad \text{with $q+\widetilde{q}$ mod $K$}\,. 
\end{equation}
Therefore, the projectors can be regarded as an order parameter that indicates crossing from a given 'universe' to another.

\section{Conclusion}

In this paper, we reviewed the construction of SymTFTs within the $AdS/CFT$ framework, emphasizing the string-theoretic origins of charged defects and symmetry topological operators. In particular, we demonstrated that the appropriate topological action of fluxbranes—constructed from D$p$-branes and NS5-branes—precisely corresponds to the Page charges of their electromagnetic dual branes, extending our earlier results in \cite{Najjar:2024vmm}. This identification suggesting a broader applicability to other brane configurations.

We further explored the continuous symmetries of the holographic 4d $\N=1$ $SU(M)$ SYM gauge theory \cite{Klebanov:2000hb}, also referred to as the KS set-up (or solution). Our primary focus was to systematically identify and analyse all potential continuous $(-1)$-form and 2-form symmetries of the KS set-up. We explicitly derived the corresponding symmetry topological operators and constructed the associated 5-dimensional SymTFT terms that encode these continuous $m$-form symmetries.

In the construction of BF-terms for continuous symmetries, one natural starting point involves reducing the topological limit of the Type IIB $p$-form kinetic terms along the cohomology basis of the link space. However, such a reduction alone generally fail to produce the correct BF-terms associated with continuous symmetries. To resolve this discrepancy, we supplemented these kinetic terms with additional 5-dimensional contributions originating from the topological Chern–Simons terms. This strategy is motivated by our work in \cite{Najjar:2024vmm}. We demonstrated that the resulting 5d action is capable of producing the correct symmetry topological operators, matching the properly dimensionally reduced topological action of the fluxbranes used to construct the continuous $(-1)$-form and 2-form symmetries.

Moreover, by employing the aforementioned 5-dimensional SymTFT and gauging a discrete 2-form symmetry, we demonstrated that a modified instanton sum can be consistently defined for the 4d $SU(M)$ SYM theory. Our results align with previous field-theoretic findings in \cite{Seiberg:2010qd,Tanizaki:2019rbk} and the M-theoretic results in \cite{Najjar:2024vmm}. Additionally, we showed that by interpreting a given 4-form field strength within the SymTFT as a 4-form gauge field of a gauged discrete 3-form symmetry on the SymTFT's physical boundary, a 4-group structure can be defined for the 4d SYM theory. This conclusion is consistent with the field-theoretic analysis presented in \cite{Tanizaki:2019rbk} and the M-theoretic results in \cite{Najjar:2024vmm}.

\acknowledgments
We thank Yi-Nan Wang and Yi Zhang for useful discussions. The author is supported by National Natural Science Foundation of China under Grant No. 12175004, No. 12422503 and by Young Elite Scientists Sponsorship Program by CAST (2023QNRC001, 2024QNRC001).


\bibliographystyle{JHEP}
\bibliography{ref}

\end{document}